\documentclass[%
 reprint,
 amsmath,amssymb,
aps,
prl,
]{revtex4-2}

\usepackage{graphicx}
\usepackage{amsmath}
\usepackage{physics}
\usepackage{amssymb}
\usepackage{xcolor}

\usepackage{tabularx}

\usepackage{booktabs}

\def\n{n}

\def\sss{\scriptscriptstyle\rm}

\def\g{_\gamma}

\def\1var{(\bx_1...\bx\N)}

\def\half{\frac{1}{2}}

\def\br{{\bf r}}

\def\bx{{x}}

\def\x{_{\sss X}}
\def\X{_{\sss X}}
\def\c{_{\sss C}}
\def\C{_{\sss C}}
\def\s{_{\sss S}}
\def\xc{_{\sss XC}}
\def\XC{_{\sss XC}}

\def\N{_{\sss N}}

\def\LO{_{\rm LO}}

\def\hyb{^{\rm hyb}}

\def\LDA{^{\rm LDA}}

\def\VW{^{\rm VW}}

\def\unif{^{\rm unif}}

\def\ee{_{\rm ee}}

\def\up{_\uparrow}
\def\dn{_\downarrow}

\def\sph_int{ {\int d^3 r}}

\def\intr{\int d^3r\,}

\graphicspath{{figures/}}

\newcommand{\rp}[1]{\textcolor{red}{#1}}

\newcounter{edit}

\renewcommand{\rp}[1]{#1}

\usepackage{pifont}
\newcommand{\cmark}{\text{\ding{51}}}
\newcommand{\xmark}{\text{\ding{55}}}

\begin{document}

%Put the title of your paper here
\title{The difference between molecules and materials: Reassessing the role of
exact conditions in density functional theory}

%Put the names of authors here:
\author{Ryan Pederson}
\email{pedersor@uci.edu}
\affiliation{Department of Physics and Astronomy, University of California, Irvine, CA 92697, USA}

\author{Kieron Burke}
\email{kieron@uci.edu}
\affiliation{Department of Chemistry, University of California, Irvine, CA 92697, USA}
\affiliation{Department of Physics and Astronomy, University of California, Irvine, CA 92697, USA}

%\date{\today}

\begin{abstract} 
Exact conditions have long been used to guide the construction of density functional approximations. 
But hundreds of empirical-based approximations tailored for chemistry are in use, many of which neglect these conditions in their design.
We analyze well-known conditions and revive several obscure ones. Two crucial distinctions are drawn: that between necessary and sufficient conditions, and between all electronic densities and the subset of realistic Coulombic ground states. Simple search algorithms find that many empirical approximations satisfy many exact conditions for realistic densities and non-empirical approximations satisfy even more conditions than those enforced in their construction. The role of exact conditions in developing approximations is revisited.

\end{abstract}
\maketitle

%\section{Introduction}

Modern density functional theory (DFT) calculations span many branches of the science of matter~\cite{norskov2011density,jain2016computational,pickard2020superconducting,zeng2019growth}. In the standard Kohn-Sham approach~\cite{KS65}, only the exchange-correlation (XC) energy need be approximated as a functional of the electronic (spin)-densities. Currently, hundreds of distinct XC approximations are available in standard DFT codes~\cite{mardirossian2017thirty,LSOM18}, reflecting the immense difficulty in finding approximations that are generally accurate.

However, there is a huge divide between the materials and molecular electronic structure
communities, as they typically use two different classes of density functional approximations.
Most materials calculations use \textit{non-empirical} semilocal approximations (generalized gradient approximations (GGAs) or meta-GGAs),
designed using \textit{exact conditions} (known analytical properties of the exact functional), as most developers and users in materials research believe that their enforcement improves performance~\cite{kaplan2022predictive}.
Such non-empirical functionals rely heavily on such conditions and eschew fitting to any chemical bonds. The recent ``Strongly Constrained and Appropriately Normed'' (SCAN)~\cite{scan_functional} semilocal functional attributes much of its success to the satisfaction of ``all known'' ($17$) exact conditions that such a functional can satisfy. 

Conversely, many \textit{empirical} approximations tailored for molecular chemistry applications {blatantly} ignore exact conditions in their design~\cite{peverati2014quest}. Such approximations can be extremely accurate on molecular benchmarks~\cite{gmtkn55}, often surpassing their more constrained counterparts.  Typically, {such chemically trained functionals} behave poorly for materials where they are seldom used.

\begin{figure}
\centering
\includegraphics[width=0.9\columnwidth]{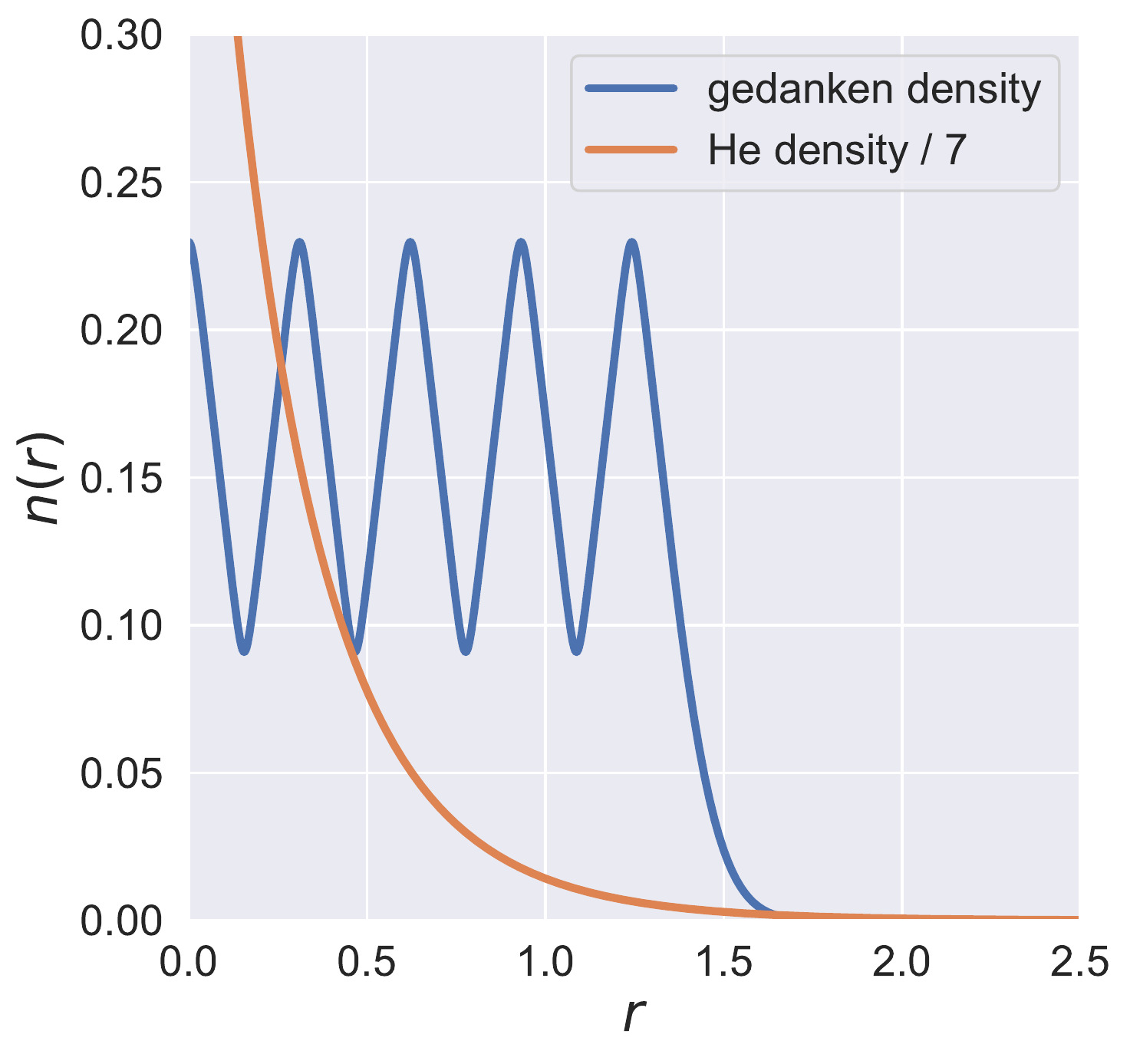}
\caption{An unpolarized two-electron ground-state gedanken density whose correlation energy is -21 mH in PBE, but +85 mH in LYP. For reference, the He atom density (divided by $7$) is plotted.}
\label{fig: gedanken density}
\end{figure}

Attaining high-accuracy for \textit{both} molecules and materials, e.g., for catalysis applications, is a major challenge due to these seemingly disjoint design paradigms.
We illustrate this difference with the correlation energy of {the blue} two-electron density in Fig.~\ref{fig: gedanken density}, calculated with two GGAs.  The first, the Perdew-Burke-Ernzerhof (PBE) correlation functional~\cite{PBE96} {adheres to many exact conditions and} automatically satisfies the basic requirement that the correlation energy is never positive, yielding
-21 mH. The second, the Lee-Yang-Parr (LYP) correlation functional~\cite{lyp1988}, does not explicitly enforce many exact conditions {and} yields the nonsensical  +85 mH.  Yet LYP has been used successfully in over 100,000 chemical applications~\cite{lyp1988}.

Why are exact conditions so important for approximations in materials, and so often ignored in molecular applications? We resolve this paradox by reassessing the role of exact conditions in modern DFT approximations.  To do this, we develop
several new (and not so new) tools.  We carefully parse the logic of exact conditions, finding that many enforced conditions are too strong for real matter.  A computational scanning procedure finds violations, coupled with construction of corresponding reasonable (but not realistic) densities, as in Fig.~\ref{fig: gedanken density}. Half a dozen exact conditions and hundreds of approximate functionals are analyzed.  Several {obscure} conditions are revived and analyzed, while even well-known conditions yield surprising new twists. Finally, the role of exact
conditions in density functional development is revisited.

Begin with the correlation energy.
In practice, approximations (denoted by tilde) have the form:
\begin{equation}
\begin{aligned}
\tilde{E}\C[n] &= \intr n(\br) \, \tilde{\epsilon}\C[n](\br),
\end{aligned}
\label{eq: def eps_c and F_c}
\end{equation}
where $n(\br)$ is an electronic density. While the developers define a conventional correlation energy per electron, $\tilde{\epsilon}\C[n](\br)$, that is often implemented explicitly in DFT codes, other ``gauges'' exist yielding the same $\tilde{E}\C[n]$. For example, $\tilde{\epsilon}\C[n](\br)$ and $\tilde{\epsilon}\C[n](\br) + \nabla^2(n^{2/3})/n$ yield identical $\tilde{E}\C[n]$~\cite{burke2014gedanken,cruz1998exchange}. 
We focus on semilocal functionals, which dominate practical calculations and are written
\begin{equation}
    \tilde{\epsilon}\C[n](\br) = \tilde{\epsilon}\C(r_s(\br), \zeta(\br), s(\br), \alpha(\br), q(\br)) \, ,
\label{eq: semilocal eps}
\end{equation}
where $r_s = (4\pi n /3)^{-1/3}$ is the Wigner-Seitz radius, $\zeta = (n_{\up} - n_{\dn}) / n$ is the (dimensionless) spin polarization, $s = |\nabla n|/(2 (3 \pi^2)^{1/3} n^{4/3})$ is the (dimensionless) reduced gradient, $\alpha = (\tau - \tau\VW)/\tau\unif {\geq 0}$  with $\tau = \sum_{i, \sigma}^{\text{occ.}} |\nabla \phi_{i, \sigma}|^2 / 2$, $\tau\VW = |\nabla n|^2 / 8n$, $\tau\unif = (3/20)(3\pi^2)^{2/3} n^{5/3} \big[ (1+\zeta)^{5/3} + (1-\zeta)^{5/3} \big]$, and $ q = \nabla^2 n / (4(3 \pi^2)^{2/3} \, n^{5/3})$ is the reduced Laplacian. 
 The local spin density approximations (LDA) depends only on $r_s$ and $\zeta$, GGAs add dependence on $s$, while meta-GGAs can depend on all variables.

{A simple condition} is \textit{correlation energy non-positivity},
\begin{equation}
    E\C[n] \leq 0 \, ,
\label{eq: E_c non-positivity}
\end{equation}
which holds for any reasonable density, which we define as being positive, integrating to a finite quantity $N$, and have finite von Weizs\"{a}cker kinetic energy ($\mathcal{I}_N$ of Ref.~\cite{L83} or Eq.~34 of Ref.~\cite{WBSB14}). 
This is routinely enforced via
\begin{equation}
    \tilde{\epsilon}\C[n](\br) \leq 0 \quad \text{for all} \, \,  \br \text{ and any } n(\br).
\label{eq: eps_c and F_c negativity condition}
\end{equation}
Clearly, satisfying {this} local condition {\em guarantees} Eq.~\eqref{eq: E_c non-positivity}, but it is also excessive, i.e., not necessary.
Moreover, starting from any $\tilde{\epsilon}\C[n](\br)$ that satisfies Eq ~\eqref{eq: eps_c and F_c negativity condition}, addition of $C \, \nabla^2(n^{2/3})/n$ violates it for sufficiently large $C$.  
If local violations of Eq.~\eqref{eq: eps_c and F_c negativity condition} {\em do} exist, then a counterexample density that violates the exact condition in Eq.~\eqref{eq: E_c non-positivity} {\em might} {be found}. If it {can be}, the exact condition is violated {for that density} in any gauge. {If no such counterexample can be found, the possibility that a gauge might be found that satisfies Eq.~\eqref{eq: eps_c and F_c negativity condition} remains open.}
    
Returning to the LYP GGA, we found instances where $\epsilon\C^{\text{LYP}}(r_s, \zeta, s) > 0$ for $s \geq 1.74$, thus allowing
the possibility of a violating \textit{gedanken} density.   Gedanken densities are thought experiment densities that need not be realistic~\cite{burke2014gedanken}. So we construct a gedanken density that has large $s \geq 1.74$ values throughout its interior, violating the local condition in Eq.~\eqref{eq: eps_c and F_c negativity condition} {(Appendix~\ref{app:1})}. Importantly, we want such local violations to exist in \textit{energetically relevant regions} of the density, that is, spatial regions that substantially contribute to the integral in Eq.~\ref{eq: def eps_c and F_c}. The gedanken density of Fig.~\ref{fig: gedanken density} is radial, nodeless, finite, continuous, {differentiable to first and second order}, and integrates to $2$ electrons. It is a reasonable density and is also {non-interacting} $v$-representable, and when evaluated using the LYP correlation functional yields +85 mH. {Thus,} LYP \textit{can} violate correlation non-positivity.
    
But does LYP violate {correlation non-positivity} in the restricted space of {realistic Coulombic densities, i.e, those ground-state densities of systems with Coulombic attractions to integer nuclear charges of small or no overall charge}? The gedanken density of Fig.~\ref{fig: gedanken density} {is not} Coulombic: for instance, it lacks nuclear cusps as required by Kato's theorem~\cite{kato1957eigenfunctions}.

\begin{figure}
\centering
\includegraphics[width=0.9\columnwidth]{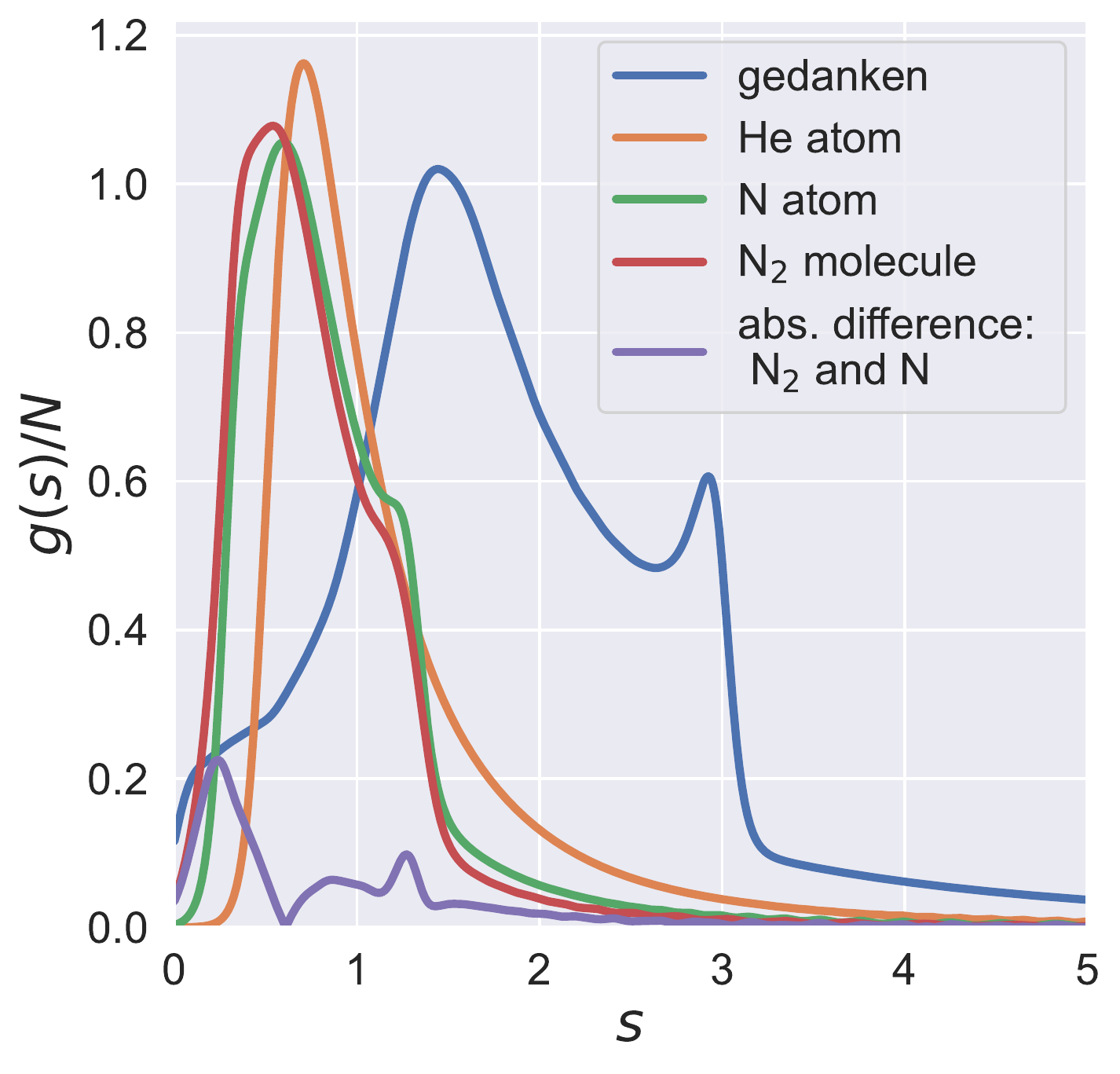}

\caption{The distribution $g(s)/N$ for various ground-state densities: the gedanken density in Fig.~\ref{fig: gedanken density}, the He and N atoms, and the N$_2$ molecule. The absolute difference between the N$_2$ molecule and N atom distributions is also plotted.}

\label{fig: g_s distribution}
\end{figure}
    
{An important property} of the gedanken density is that it has energetically relevant regions with $s \geq 2$. {The distribution}
\begin{equation}
    g(s) = \intr n(\br) \, \delta^{(3)}(s - s(\br)) \, ,
\end{equation}
was introduced in Refs.~\cite{zupan1997density,zupan1997distributions}, {and} $g(s) \, ds$ is the number of electrons in the system with reduced density gradient between $s$ and $s+ds$, i.e, it is an analog of the density of states for energy levels. In Fig.~\ref{fig: g_s distribution}, we plot $g(s)/N$ for various ground-state densities: the gedanken density in Fig.~\ref{fig: gedanken density} and {calculated} densities for the He and N atoms and the N$_2$ molecule. Hartree-Fock (HF) densities {are sufficiently accurate for our purposes}. Unlike the Coulombic densities, the gedanken density $g(s)$ {is centered} around $s \approx 2$, {as intended}. 
For Coulombic systems~\cite{zupan1997distributions}, large $s > 2$ values are typically only found in the decaying tails of the density, and are energetically irrelevant. In molecular and extended systems, these tails (which may not even be present in periodic systems) are even less energetically relevant than their atomized counterparts~\cite{zupan1997distributions}: in Fig.~\ref{fig: g_s distribution} the distribution $g(s)/N$ of N$_2$ is {shifted to lower $s$} than that of the atomized system (N atom) for $s > 1$. 
The electrons in the bond between two (or more) atoms have {smaller values of $s$ which even vanishes} at the bond center~\cite{zupan1997distributions}. {Although a single electron distributed across an infinitely separated chain of many protons has energetically relevant regions of large $s$, such a system is too far from neutral for our set.}

{Do} large $s$ values contribute importantly to energy differences, such as molecular binding energies? {Valence} electrons change considerably but their differences tend to be energetically relevant in regions of smaller $s$ values {: the difference} $|g[\text{N}_2](s) - 2g[\text{N}](s)|$, {is prominent} for $s < 2$ (Fig.~\ref{fig: g_s distribution})~\cite{zupan1997distributions}. In non-covalent bonding, $s$ values up to $ \approx 7$ {are} relevant in binding energy differences for van der Waals (vdW) complexes~\cite{jenkins2021reduced,murray2009investigation}, but typically a non-semilocal correlation functional (such as DFT-D~\cite{grimme2010consistent} or vdW-DF~\cite{chakraborty2020next}) provides the bulk of the energetics that contributes to the binding~\cite{jenkins2021reduced}.

Revisiting LYP correlation, we conjecture \rp{that} {no} realistic Coulombic density {ever yields} a positive correlation energy.
Such a density would need to have energetically relevant regions of the density with large $s > 1.74$, which is seldom observed in these systems. \rp{Over countless} atomic and molecular densities, LYP correlation {has} \textit{not} {yielded} positive correlation energies~\cite{lyp1988,miehlich1989results}. 

We perform an identical analysis on other representative and well-known approximations and tabulate the results in the first row of Table~\ref{tab: functionals}.
{For} each exact condition (e.g. Eq.~\eqref{eq: E_c non-positivity}), we check for violations of the corresponding local condition (e.g. Eq.~\eqref{eq: eps_c and F_c negativity condition}) for various semilocal approximations. To locate violations, we use a simple extensive grid search over applicable variables, $0<r_s \leq 5, 0 \leq \zeta \leq 1, 0 \leq s \leq 5, 0 \leq \alpha \leq 5$, where the upper ranges are reasonably large to encapsulate relevant regions of realistic Coulombic densities. This brute-force numerical approach is necessary since analytical verification is often impossible {with} the complicated forms of modern XC approximations. 
\rp{We} utilize the Libxc~\cite{LSOM18} implementation of XC approximations to evaluate the XC energy densities in Eq.~\eqref{eq: semilocal eps}.

For a given local condition and approximation, if we find no violations in our grid search, then we \rp{assume} that the approximation always satisfies the exact condition \rp{for any reasonable density (denoted by a \cmark \, in Table~\ref{tab: functionals})}. 
In other cases {(denoted \cmark *)}, the corresponding local condition may be satisfied only for a restricted range of variable values. 
For example, in B3LYP the local condition in Eq.~\ref{eq: eps_c and F_c negativity condition} is satisfied whenever $s < 2.13$ {and we display the bounds on $s$ that ensure satisfaction}. 
In some cases {(denoted \cmark **)}, we do not obtain a simple restricted range of variable values, but we find that local violations are exceedingly ``rare'' \rp{(less than $1\%$ of the total number of configurations considered)}. 

\setlength{\tabcolsep}{2pt}
\begin{table*}[ht]
\centering
\renewcommand{\arraystretch}{2}
\begin{tabular}{|c|c|c|c|c|c|c|c|c|}
\hline
& \multicolumn{3}{|c|}{non-empirical} & \multicolumn{5}{|c|}{empirical}\\
\hline
local condition & PBE & AM05 & SCAN & B3LYP & CASE21 & SOGGA11 & M06 & B97
\\ \hline
$E\C$ non-positivity~\eqref{eq: eps_c and F_c negativity condition} & \cmark & \cmark & \cmark & \cmark *, $s < 2.13$ & \cmark & \cmark & \cmark *, $\zeta = 0, s< 1.56$ & \cmark *, $s < 1.42$ 
\\ \hline
$E\C$ scaling inequality~\eqref{eq:Fc lower bd cond} & \cmark & \cmark & \cmark & \cmark *, $s < 2.15$ & \cmark & \cmark ** & \cmark *, $\zeta = 0, s < 1.59$ & \cmark *, $s < 1.52$ 
\\ \hline
$T\C$ upper bound~\eqref{eq:Fc upper bound cond 1} & \cmark & \cmark & \cmark & \cmark & \cmark & \cmark *, $s < 1.36$ & \cmark *, $\zeta = 0, s < 1.56$ & \cmark *, $s < 1.62$ \\ \hline
\rp{$U\C(\lambda)$ monotonicity} ~\eqref{eq:Fc second deriv cond} & \cmark & \cmark & \cmark & \cmark *, $s < 1.82$ & \cmark & \xmark & \cmark *, $\zeta = 0, s < 1.56$ & \cmark *, $s < 1.41$ 
\\ \hline
LO extension to $E\xc$~\eqref{eq: Fxc LO Exc} & \cmark & \cmark & \cmark & \cmark & \cmark & \cmark & \cmark *, $0.04 < s < 3.62$ & \cmark *, $s < 4.46$
\\ \hline
LO ~\eqref{eq: Fxc LO Uxc} & \cmark & \cmark & \cmark & \cmark *, $s < 4.88$  & \cmark & \cmark *, $s < 4.98$ & \cmark *, $0.06 < s < 3.62$ & \cmark *, $s < 4.43$
%\\ \hline \hline
%Conjecture: $T_c \leq -E_c$ ~\eqref{eq:Fc upper bound cond 2} & \cmark ** & \cmark & \cmark & \cmark *, $s<1.64$ & \xmark & \cmark ** & \xmark & \cmark *, $s<1.43$
\\ \hline 
\end{tabular}
\parbox{\textwidth}{\caption{For each condition, we assess if the local condition is satisfied (or partially satisfied) for an approximation {(with more given in Supplemental S1).}}
\label{tab: functionals}}
\end{table*}

%\section{exact conditions}

The logic and concepts presented {for} correlation non-positivity generalize to other exact conditions on energy functionals. 
For instance, the \textit{Lieb-Oxford (LO) bound extension to XC energies exact condition}~\cite{LO81,pw91}, 
\begin{equation} 
    E\XC[n] \geq C\LO \intr n \, \epsilon_{\X}\unif[n](\br) \, ,
\label{eq:LO Exc}
\end{equation}
{yields the} local condition
\begin{equation}
    \tilde{F}\XC \leq C\LO \, ,
\label{eq: Fxc LO Exc}
\end{equation}
where we use $C\LO = 2.27$ in this work (although {tighter} bounds have been proven ~\cite{LLS19,LLS22}), $\tilde{F}_{\sss X(C)}[n](\br) \equiv \tilde{\epsilon}_{\sss X(C)}[n](\br) / \epsilon\x^{\text{unif}}[n](\br)$ is the {exchange (correlation) enhancement factor with $\tilde{F}\XC = \tilde{F}\X + \tilde{F}\C$ and} $\epsilon\x^{\text{unif}}[n](\br) = - (3/4 \pi)(3 \pi^2 n)^{1/3}$ is the exchange energy per particle of {an unpolarized} uniform electron gas. 
Many approximations \rp{enforce the local Eq.~\eqref{eq: Fxc LO Exc} to ensure Eq.~\ref{eq:LO Exc}}. 

Since \rp{the combined XC energy is the object of interest, some} approximations {fail to distinguish} exchange and correlation. In the exact functional, one can {extract $E\c$ using} uniform coordinate scaling~\cite{levy1985hellmann,levy1993tight}:
\begin{equation}
    E\c[n] = E\xc[n] - \lim_{\gamma \to \infty} \frac{E\xc[n\g]}{\gamma} \, ,
\label{eq:conventional partitioning}
\end{equation}
where $n\g(\br) \equiv \gamma^3 n(\gamma \br)$ and $\gamma > 0$. We apply this ``conventional'' partitioning to extract correlation energies {where none have been defined} {or whose partitioning is ambiguous. E.g., for global hybrids, Eq.~\eqref{eq:conventional partitioning} {yields $E\c$ of} the semilocal form in Eq.~\eqref{eq: semilocal eps}, e.g., $\epsilon\c^{\text{B3LYP}} = 0.405 \epsilon\c^{\text{LYP}} +  0.095 \epsilon\c^{\text{VWN5}}$~\cite{stephens1994ab}. {This} partitioning {can} differ from the developer's intentions or rationalizations (Appendix~\ref{app:4}). The LO bound applied to global hybrid functionals is discussed in Appendix~\ref{app:5}.}

Besides Eq.~\eqref{eq:conventional partitioning}, many other properties of the exact functional are written in terms of uniform coordinate scaling (or equivalently, through the adiabatic connection in DFT~\cite{HJ74, LP75, GL76}, see Appendix~\ref{app:7}). \rp{We} simply list these {``obscure''} conditions and their \rp{local forms} {(with details and derivations in {Appendix~\ref{app:3}})}. \rp{The} \textit{correlation uniform scaling inequality}~\cite{levy1985hellmann}
\begin{equation}
    (\gamma - 1) E\C[n\g] \geq \gamma (\gamma - 1) \, E\C[n] 
\label{eq: correlation uniform scaling inequality}
\end{equation}
\rp{has a} corresponding local condition,
\begin{equation}
    \frac{\partial \tilde{F}\c(r_s, \zeta, s, \alpha, q)}{\partial r_s} \geq 0 \, .
\label{eq:Fc lower bd cond}
\end{equation}
The kinetic contribution to the correlation energy, $T\C$, is {non-negative}~\cite{levy1985hellmann,frydel2000} 
\begin{equation}
    T\c[n\g] = \gamma \frac{d E\c[n\g]}{d\gamma} - E\c[n\g] \geq 0 \, , 
\label{eq: Tc non-negativity exact condition}
\end{equation}
and shares the same \rp{local condition, Eq.~\ref{eq:Fc lower bd cond}}. 
The $T\c$ {\textit{upper bound}}~\cite{levy1994density,levy1991density} reads
\begin{equation}
    T\c[n\g] \leq  -\gamma \bigg(\frac{\partial E\c[n\g]}{\partial \gamma} \Bigr|_{\gamma \to 0} \bigg) + E\c[n\g] \, ,
\label{eq:Tc upperbound exact cond}
\end{equation}
with corresponding local condition,
\begin{equation}
    \frac{\partial \tilde{F}\c}{\partial r_s} \leq \frac{\tilde{F}\c(\infty) - \tilde{F}\c}{r_s} \, ,
\label{eq:Fc upper bound cond 1}
\end{equation}
\rp{where $\tilde{F}\c(\infty) = \tilde{F}\c(r_s \to \infty)$.}
\rp{Correlation energy adiabatic connection curves, $U\C(\lambda) = d(\lambda^2 E\C[n_{1/\lambda}])/d\lambda$, satisfy a \textit{monotonicity condition}~\cite{levy1993tight},
\begin{equation}
    \frac{d U\C(\lambda)}{d \lambda}  \leq 0 \, ,
\end{equation}
}
with corresponding local condition~\cite{levy1993tight}
\begin{equation}
    \frac{\partial}{\partial r_s} \bigg(r_s^2 \frac{\partial \tilde{F}\c}{\partial r_s} \bigg) \geq 0 \, .
\label{eq:Fc second deriv cond}
\end{equation}
The LO bound~\cite{LO81}, {often generalized as Eq.~\ref{eq:LO Exc}, is precisely}
\begin{equation} 
    U\XC[n] \geq C\LO \intr n \, \epsilon_{\X}\unif[n](\br) \, ,
\label{eq:LO Uxc}
\end{equation}
{where $U\XC[n] = E\XC[n] - T\C[n]$ is the potential correlation energy. The} corresponding local condition,
\begin{equation}
    \tilde{F}\XC + r_s \frac{\partial \tilde{F}\c}{\partial r_s} \leq C\LO \, ,
\label{eq: Fxc LO Uxc}
\end{equation}
{is \textit{more}} restrictive than the commonly used Eq.~\eqref{eq: Fxc LO Exc}. 
{Approximations satisfying Eq.~\eqref{eq: Fxc LO Exc} need not satisfy Eq.~\eqref{eq: Fxc LO Uxc}, such as B3LYP or SOGGA11 in Table~\ref{tab: functionals}}. Results for a conjectured condition, $ T\C[n] \leq -E\C[n]$ ~\cite{levy1985hellmann,crisostomo2022seven,frydel2000}, {are} in Appendices~\ref{app:6} and ~\ref{app:10} and Supplemental S1. 

{This work does not provide a comprehensive study of all known exact conditions in DFT. A unified subset of several conditions (6) were chosen to illustrate the logic.} Other well-known deficiencies of semilocal approximations include the self-interaction error~\cite{PZ81}, the asymptotic behavior of exchange and correlation potentials~\cite{almbladh1985exact}, or the flat-plane {energy condition for} fractional charges and spins~\cite{MCY09,CMW12}. 
\rp{We} expect our {logic} can be fruitfully applied to any exact condition in DFT.

We describe the \rp{conditions} Eqs.~\ref{eq: correlation uniform scaling inequality} - ~\ref{eq: Fxc LO Uxc} as obscure because, while proven several decades ago, none are deliberately and generally enforced in modern approximations, even \rp{those} that strive to satisfy many exact conditions. {SCAN was designed to satisfy the correlation uniform scaling inequality (Eq.~\eqref{eq: correlation uniform scaling inequality}), but only in extreme limits, $\gamma \to 0, \infty$~\cite{scan_functional,kaplan2022predictive}.} {The} corresponding local condition in Eq.~\eqref{eq:Fc lower bd cond} \textit{is} satisfied in {SCAN, but adjustments of its parameters chosen to fit \textit{appropriate norms} can produce violations (Appendix~\ref{app:9}). Norms} refer to properties of specific (but not bonded) reference systems, such as the uniform gas, the hydrogen atom, or noble gas \rp{dimers}~\cite{SRP15,becke2022density}.
{So, by enforcing appropriate norms, SCAN satisfies \textit{more} exact conditions than were explicitly included}.
{Similarly, LDA has long been known to inherit exact conditions from its uniform gas norm~\cite{LP75,GL76,levy1985hellmann,levy1993tight}. We find that many parameterizations of LDA correlation, such as PW92~\cite{perdew1992accurate} and VWN~\cite{vosko1980accurate}, also satisfy these obscure exact conditions.}
Non-empirical approximations PBE, AM05, and SCAN \rp{functionals}, which were designed to satisfy a large set of exact conditions, also satisfy many additional exact conditions outside of the original set (Table~\ref{tab: functionals}). CASE21, a recent machine-learned empirical functional \rp{designed} to adhere to many exact conditions~\cite{distasio2021case21}, \rp{also satisfies} these obscure conditions.

In Table~\ref{tab: functionals}, we find many empirical approximations satisfy local conditions in energetically relevant regions of {realistic} Coulombic densities, i.e., for moderate $s$ values. 
When assessing exact conditions on coordinate-scaled HF densities for atoms H-Ar and their cations, we find \rp{that} \textit{all} are satisfied (Appendix~\ref{app:10}), suggesting that these approximations will satisfy these conditions in the space of {realistic} Coulombic densities ({possibly excepting the monotonicity condition in SOGGA11}). 
This is intriguing because {most} such empirical approximations were designed \rp{\textit{without}} explicit adherence to these exact conditions. 
This finding appears to reinforce their importance in approximations: satisfaction of such esoteric conditions is {hardly accidental}.
Furthermore, empirical approximations often employ ingredients, such as the dimensionless quantities $s, \zeta, \alpha, q$, which themselves were chosen to simplify satisfaction of exact conditions. 
In consequence, nearly all empirical approximations satisfy two simple exact conditions on the exchange energy: uniform coordinate scaling~\cite{levy1985hellmann} and spin scaling~\cite{OP79}.

{Our results suggest} a reassessment of the role of exact conditions in modern density functional development. 
{Is it excessive to enforce strong local conditions to ensure the satisfaction of exact conditions?
Flexible empirical model approximations also satisfy many exact conditions on the (highly relevant) space of {realistic} Coulombic densities and achieve high accuracy for molecular processes.
However, when empirical functionals are locally constrained to satisfy many conditions, their molecular benchmark performance \rp{is similar to the suboptimal performance of their non-empirical counterparts},
despite the advantage of training on molecular data~\cite{distasio2021case21,dick2021highly,nagai2022machine}\rp{, and} the resulting functionals tend to closely mimic non-empirical counterparts, such as SCAN~\cite{dick2021highly,nagai2022machine}.}

So when is the enforcement of local conditions helpful?  The answer appears to lie in the paucity of high-accuracy data for solids.  Empirical approximations are mostly developed for molecules, where copious \rp{benchmark} data is now available. Alternatively, non-empirical functionals that dominate materials calculations include norms such as the uniform gas limit, which substitute for high-accuracy data.  By reducing to this limit, non-empirical functionals are guaranteed to yield moderately accurate results for solids.  Moreover, the leading corrections in the asymptotic limit differ qualitatively between molecules and solids, because all molecules have turning surfaces at the Kohn-Sham HOMO energy, while few solids do~\cite{kaplan2021calculation}.  This produces conflicting requirements on the gradient expansion of the approximation, as shown in the differences between PBE (good for atomic and molecular energies) and PBEsol (good for solid geometries and vibrations~\cite{PRCV08}).   Such conflicts {are} resolved in SCAN, yielding improved results for both. {Even in the presence of high-accuracy data for solids, approximations that fit their data will worsen their results for molecules.}

{Even for molecules, non-empirical approximations may also outperform empirical counterparts, especially for systems or properties outside their training~\cite{kaplan2022predictive,medvedev2017density,kanungo2021comparison}.
For instance, for the artificial molecules in the MB16 benchmark~\cite{korth2009mindless,gmtkn55} (never used to parameterize empirical functionals), SCAN tends to outperform empirical approximations, including hybrids~\cite{kaplan2022predictive}, suggesting a greater ability to extrapolate.}

This work resolves a central tension of DFT development: Why are exact conditions so important for DFT in materials, and so often ignored in molecular applications?  In molecules, vast reliable databases implicitly enforce relevant conditions on realistic densities.  But the dearth of such databases for materials is overcome by the enforcement of conditions on all possible densities, ensuring reasonable (if not highly accurate) results for materials.  

Why is this important?  First, we have shown that enforcing conditions locally (to guarantee satisfaction) is typically excessive.  Second, we have found that even forgotten conditions are often automatically satisfied by non-empirical functionals, suggesting consistency in their design.  
Lastly, this work suggests that imposition of \textit{select} exact conditions (and appropriate norms) can substitute for materials reference data and maintain the transferability of resulting approximations across molecules, solids, and everything in between, e.g., interfaces and clusters.

\section*{Acknowledgments}
Work supported by DOE DE-SC0008696 and the Eddleman Quantum Institute. We thank John P. Perdew and Aaron D. Kaplan for useful discussions.

\section*{Data availability}

The data and analysis code that supports the findings of this study are available at \url{https://github.com/pedersor/DFT_exconditions}, with further details provided in the appendices and supplementary material.

% word count
% 2836 + 17*16 + 170*2 + 26+(13*8)
% = 3578
% https://journals.aps.org/authors/length-guide
% max = 3750

\appendix

\section{Gedanken density details}
\label{app:1}

\begin{figure}[h]
\centering
\includegraphics[width=0.40\textwidth]{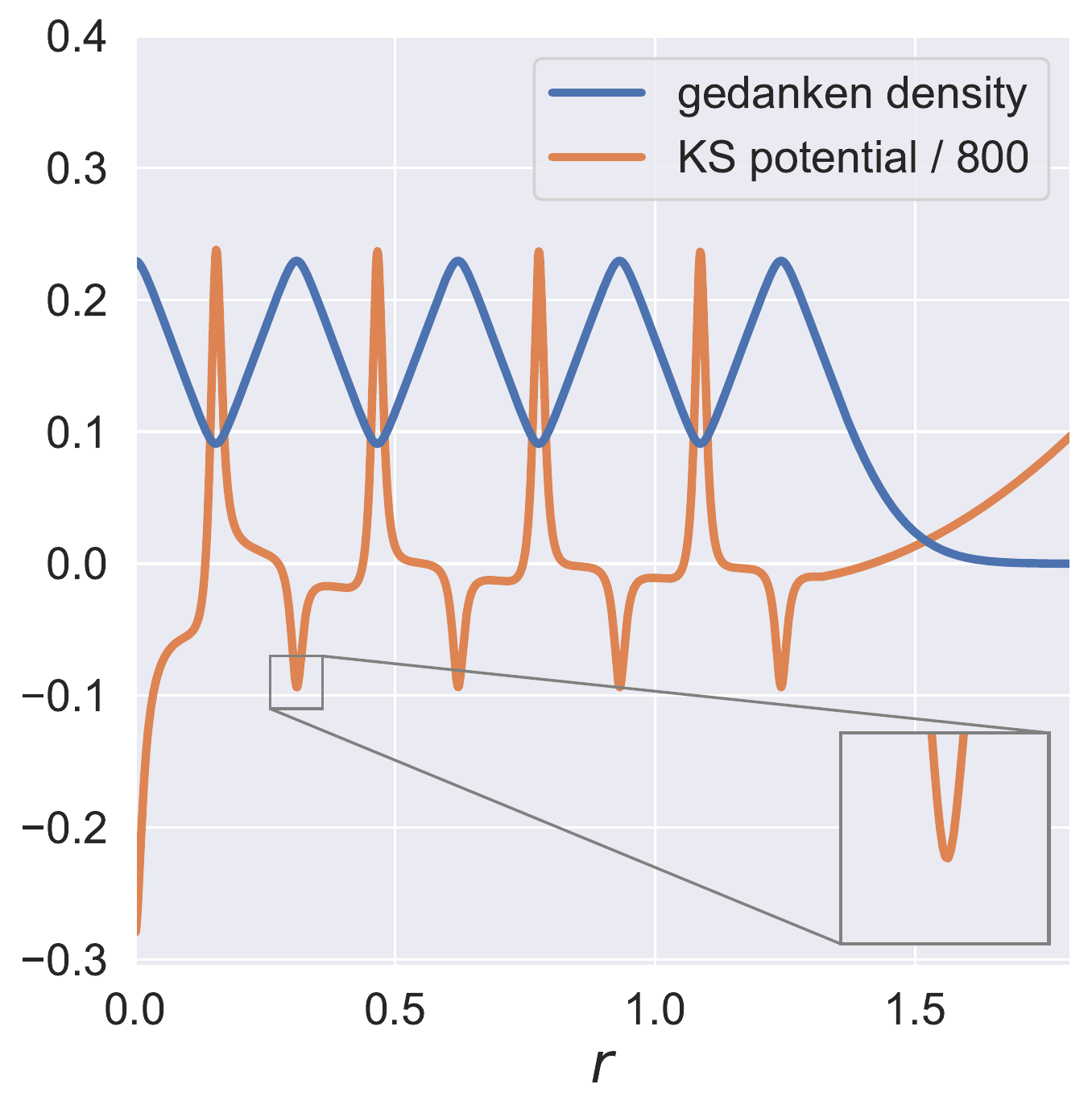}
\caption{The example Gedanken density considered in the main text and its corresponding KS potential. The potential is continuous everywhere.}
\label{fig: gedanken density and potential}
\end{figure}

The Gedanken density used in the main text has the radial form
{
\begin{equation}
n(r) \propto \begin{cases} 
C - \frac{A}{\pi}\cos^{-1} \bigg((1-\eta) \sin( \frac{2 \pi r}{T} - \frac{\pi}{2})\bigg) & r \leq r_p \\
e^{-a r^2 + b r + c} & r > r_p 
\end{cases}
\end{equation}
}
where the density is a dampened triangle wave starting from the origin where $\eta \geq 0$ controls the smoothness of the waveform (where $\eta = 0$ produces a triangle wave), $T > 0$ is the period of the waveform, $N_p \geq 0$ is an integer that controls the number of peaks in the waveform, $A > 0$ is the amplitude, and $C>0$ is an offset. At $r_p = (N_p - 3/4)T$, the density transitions to a Gaussian, where $a$, $b$, and $c$ are determined by ensuring continuity and first and second derivative continuity. The final gedanken density is then normalized to the desired number of electrons ($2$ here). {Specific variable values used for the gedanken density in the main text are provided in Table~\ref{tab:gedanken variables}.} In Fig.~\ref{fig: gedanken density and potential} we plot
this gedanken density and its corresponding Kohn-Sham (KS) potential,
{
\begin{equation}
    v\s(r) = \half \frac{ \frac{d^2}{dr^2} \big(r \sqrt{n(r)}\big)}{r \sqrt{n(r)}} \, .
\end{equation}
}

\begin{table}[h]
\label{tab:gedanken variables}
\begin{tabular}{|c|c|}
\toprule
$C$ & 0.2387324146 \\
$A$ & 0.1679968844 \\
$\eta$ & 0.05 \\
$T$ & 0.3105085788 \\
$N_p$ & 5 \\
$a$ & 22.0154308155 \\
$b$ & 51.4622187780 \\
$c$ & $2.2 \times 10^{-14}$ \\
\bottomrule
\end{tabular}
\caption{Variable values used in the example gedanken density.}
\end{table}

{In Table~\ref{tab:exchange gedanken}, the exchange energy for the gedanken density is given for different exchange approximations, including B88 exchange~\cite{becke1988density}, which was explicitly designed to the correct large $r$ behavior for Coulombic systems.
Two-electron densities, such as the gedanken density, follow a tight Lieb-Oxford (LO) bound for exchange, $E\X[n] \geq 1.174 \, E\LDA\X[n]$~\cite{burke2014gedanken}. For the gedanken density, $1.174 \, E\LDA\X = -0.925$ and we see that PBE and B88 exchange violate this exact condition. 
On the other hand, SCAN explicitly enforces this tight bound condition and does not violate it. }

\begin{table}[h]
\begin{tabular}{|c|c|c|c|c|}
\toprule
exact & SCAN & LDA & PBE & B88 \\
\midrule
-0.832 & -0.898 & -0.788 & -1.062 & -1.136 \\
\bottomrule
\end{tabular}
\caption{Exchange energies (in atomic units) computed on the two-electron Gedanken density. }
\label{tab:exchange gedanken}
\end{table}

\section{computational details}
\label{app:2}

All atomic calculations were performed using the PySCF code~\cite{sun2020recent} with the aug-pcseg-4 basis set. The distribution $g(s)$ was computed using a fermi distribution smoothening with fictitious temperature $T = 0.05$ (following Ref.~\cite{zupan1997density}). For some systems, increased radial grids (up to $500$) were used in the Gauss-Chebyshev grid scheme~\cite{krack1998adaptive} to maintain high fidelity of $g(s)$ at large $s$ values. Further details can be found in our public code~\cite{dft_exconditions}.

\section{Local conditions derivations and relations}
\label{app:3}

We derive several of the local conditions used in the main text which have not been previously reported in the literature to our knowledge. We start with the $T\c$ \textit{upper bound exact condition}~\cite{levy1994density} 
\begin{equation}
    T\c[n\g] \leq  -\gamma \bigg(\frac{\partial E\c[n\g]}{\partial \gamma} \Bigr|_{\gamma \to 0} \bigg) + E\c[n\g] \, ,
%\label{eq:Tc upperbound exact cond}
\end{equation}
where~\cite{frydel2000}
\begin{equation}
    T\c[n\g] = \gamma \frac{\partial E\c[n\g]}{\partial \gamma} - E\c[n\g] .
\label{eq: Tc def}
\end{equation}
For the semilocal approximations considered here,
\begin{equation}
    \tilde{E}\C[n\g] = \gamma \intr n(\br) \epsilon^{\text{unif}}\x[n](\br) \, \tilde{F}\c(r_s / \gamma, \zeta, s, \alpha, q) \, .
\label{eq: Ec DFA gamma}
\end{equation}
This is due to the fact that the dimensionless quantities $\zeta$, $s$, $\alpha$, and $q$ are all scale-invariant, i.e. $\zeta[n\g](\br) = \zeta[n](\gamma \br)$, $s[n\g](\br) = s[n](\gamma \br)$, and so on. Thus the $\gamma$-dependence in the approximation only shows up in the local quantity, $r_s[n\g](\br) = r_s[n](\gamma \br) / \gamma$. {For compactness, in the following we drop the explicit dependence on scale-invariant quantities}. 
Substituting in Eq.~\eqref{eq: Ec DFA gamma} into Eq.~\eqref{eq:Tc upperbound exact cond} and enforcing the inequality on the integrands (that is, the local enforcement the exact condition) yields
\begin{equation}
    \gamma \frac{\partial \tilde{F}\C(r_s/\gamma)}{\partial \gamma} 
    \geq -\tilde{F}\C(\infty) + \tilde{F}\C(r_s/ \gamma) \, .
\end{equation}
Using the chain rule and rearranging we obtain the following local condition
\begin{equation}
    \frac{\partial \tilde{F}\c}{\partial r_s} \leq \frac{\tilde{F}\c(\infty) - \tilde{F}\c}{r_s} \, .
%\label{eq:Fc upper bound cond 1}
\end{equation}
{The limit $\tilde{F}\c(\infty)$ is the \textit{strictly correlated electron}~\cite{seidl2007strictly} limit, which is discussed further in Section~\ref{sec:Relation to adiabatic connection}.}

The LO bound~\cite{LO81} exact condition reads
\begin{equation} 
    U\XC[n] \geq C\LO \intr n \, \epsilon_{\X}\unif[n](\br) \, ,
%\label{eq:LO Uxc}
\end{equation}
where $U\XC[n] = E\XC[n] - T\C[n]$. For the semilocal approximations considered here, utilizing Eq.~\eqref{eq: Tc def} evaluated at $\gamma = 1$ we have 
\begin{equation}
    \tilde{U}\XC[n] = \intr \n(\br) \epsilon^{\text{unif}}\x[n](\br) \bigg[ \tilde{F}\XC + r_s \frac{\partial \tilde{F}\c}{\partial r_s} \bigg] \, .
\end{equation}
Enforcing the LO bound locally yields the following local condition,
\begin{equation}
    \bigg(1 + r_s \frac{\partial}{\partial r_s} \bigg) \tilde{F}\XC \leq C\LO \, .
%\label{eq: Fxc LO Uxc}
\end{equation} 

It is also known that $T\c[n] \geq 0$, and Eq.~\eqref{eq: Tc def} yields the following condition  
\begin{equation}
    \gamma \frac{\partial E\c[n\g]}{\partial \gamma} - E\c[n\g] \geq 0 \, . 
%\label{eq: Tc non-negativity exact condition}
\end{equation}
However one can show that this condition is automatically satisfied under the \textit{correlation uniform scaling inequality} exact condition~\cite{levy1985hellmann},
\begin{equation}
\begin{aligned}
    E\C[n\g] \geq \gamma \, E\C[n]  \qquad (\gamma > 1) \, ,\\
    E\C[n\g] \leq \gamma \, E\C[n]  \qquad (\gamma < 1) \, .
\end{aligned}
%\label{eq: correlation uniform scaling inequality}
\end{equation}
Let $\gamma > 0$ be arbitrary, {$\epsilon > 0$},  with $\gamma_+ = 1 + \epsilon/\gamma$ and $\gamma_- = 1 - \epsilon/\gamma$, such that $\epsilon/\gamma << 0$ (infinitesimal). From Eq.~\eqref{eq: correlation uniform scaling inequality} we have
\begin{equation}
    E\C[n_{\gamma \gamma_+}] - E\C[n_{\gamma \gamma_-}] \geq \gamma_+ E\C[n\g] - \gamma_- E\C[n\g] \, .
\end{equation}
Simplifying yields
\begin{equation}
    \gamma \frac{E\C[n_{\gamma + \epsilon}] - E\C[n_{\gamma - \epsilon}]}{2 \epsilon} \geq E\C[n\g] \, .
\end{equation}
Applying the definition of a derivative and identifying $T\C[n\g]$ we obtain {$T\C$ \textit{non-negativity}:}
\begin{equation}
    T\C[n\g] = \gamma \frac{\partial E\C[n\g]}{\partial \gamma} - E\C[n\g] \geq 0 \, .
\end{equation}
Enforcing this condition locally yields the following local condition for approximations
\begin{equation}
    \frac{\partial \tilde{F}\c(r_s, \zeta, s, \alpha, q)}{\partial r_s} \geq 0 \, ,
%\label{eq:Fc lower bd cond}
\end{equation}
which is the same one that corresponds to Eq.~\eqref{eq: correlation uniform scaling inequality}.

{The} two inequalities in Eq.~\eqref{eq: correlation uniform scaling inequality} are equivalent. For instance, let $\gamma' < 1$ be arbitrary (but strictly positive) and take $\gamma = 1/\gamma' > 1$. Take $n \mapsto n_{\gamma'}$ in Eq.~\eqref{eq: correlation uniform scaling inequality} and we have $E\C[n_{\gamma' \gamma}] \geq \gamma E\C[n_{\gamma'}]$. Substituting for $\gamma$ and rearranging we have $E\C[n_{\gamma'}] \leq \gamma' E\C[n]$ with $\gamma' < 1$.

Finally, we note that if the local conditions in Eqs.~\eqref{eq:Fc lower bd cond} and ~\eqref{eq:Fc upper bound cond 1} are satisfied, then we have
\begin{equation}
    \tilde{F}\XC \leq \tilde{F}\XC + r_s \frac{\partial \tilde{F}\c}{\partial r_s} \leq \tilde{F}\X + \tilde{F}\c(\infty) .
\end{equation}
Since the exchange energy follows the simple scaling relation, $E\X[n\g] = \gamma E\X[n]$, $\tilde{F}\X$ needs to be scale-invariant and thus independent of $r_s$. The rightmost side of the inequality, $\tilde{F}\X + \tilde{F}\c(\infty)$, is then the maximum value that $\tilde{F}\XC$ can take (assuming all other variables besides $r_s$ are fixed). Therefore, if an approximation satisfies Eqs.~\eqref{eq:Fc lower bd cond} and ~\eqref{eq:Fc upper bound cond 1}, then the local condition 
\begin{equation}
    \tilde{F}\XC \leq C\LO \, ,
%\label{eq: Fxc LO Exc}
\end{equation}
which is the standard one that corresponds to the LO bound on $E\XC$, will imply Eq.~\eqref{eq: Fxc LO Uxc}. 
In Table~1 in the main text, indeed we see that functionals like PBE, AM05, PBE, and CASE21, which simultaneously satisfy Eqs.~\eqref{eq:Fc lower bd cond}, ~\eqref{eq:Fc upper bound cond 1}, and ~\eqref{eq: Fxc LO Exc}, automatically satisfy the LO in Eq.~\eqref{eq:LO Uxc}.

We also remark that an enhancement factor $\tilde{F}\c$ that satisfies the conditions in Eqs.\eqref{eq:Fc lower bd cond}, \eqref{eq:Fc upper bound cond 1}, and \eqref{eq: Fxc LO Exc} is monotonic and Lipschitz continuous in $r_s$, i.e., the derivative is bounded, $0 \leq \partial \tilde{F}\c / \partial r_s \leq K$, where $K$ is a finite constant known as the Lipschitz constant. Such a property of the enhancement factor may help assuage possible issues during numerical integration~\cite{lehtola2022many}. However, this property is only with respect to the $r_s$ variable and is clearly not sufficient to ensure stability.  

\section{conventional exchange and correlation partitioning}
\label{app:4}

{
We define a global hybrid functional approximation as 
\begin{equation}
    \tilde{E}\XC\hyb[n] = \tilde{E}\XC[n] + \tilde{a} E\X[n] \, ,
\end{equation}
where $\tilde{a}>0$,$E\X[n]$ is the exact exchange energy, and $\tilde{E}\XC[n]$ is the remaining semilocal density functional that can be expressed as
\begin{equation}
    \tilde{E}\XC[n] = \intr n(\br) \tilde \epsilon\XC(r_s, \zeta, s, \alpha, q) \, .
\end{equation}
Note the difference between $\alpha$ and other usual definitions for the mixing parameter in global hybrids.
}

{In hybrid XC functionals and other available approxmations, the partitioning of the exchange and correlation may be ambiguous or not defined.
In these cases, we use the following conventional partitioning
\begin{equation}
    E\c[n] = E\xc[n] - \lim_{\gamma \to \infty} \frac{E\xc[n\g]}{\gamma} \, ,
%\label{eq:conventional partitioning}
\end{equation}
which holds for the exact functional~\cite{levy1993tight}.
In global hybrid functionals, the exact exchange contributions cancel out in Eq.~\eqref{eq:conventional partitioning}, since $E\X[n\g] = \gamma E\X[n]$, and do not contribute to the correlation energy. Here the conventional partitioning for the correlation energy can be expressed as a semilocal density functional with the following correlation energy per electron
\begin{equation}
    \tilde \epsilon\C[n](\br) = \tilde \epsilon\XC(r_s, \zeta, s, \alpha, q) - \lim_{\gamma \to \infty} \frac{\tilde \epsilon\XC(r_s/\gamma, \zeta, s, \alpha, q)}{\gamma} \, .
\label{eq:eps_c conventional partitioning}
\end{equation}
}

{
In many cases, the conventional partitioning is consistent with the one defined by the authors of the approximation (if one exists). This agreement occurs whenever an approximation satisfies: $\tilde E\x[n\g] = \gamma \tilde E\x[n]$ and $\lim_{\gamma \to \infty} \tilde E\c[n\g] / \gamma \to 0$. The latter is satisfied when $\tilde E\c[n\g]$ goes to a finite constant as $\gamma \to \infty$, but also for approximations of the form
\begin{equation}
    \tilde E\C[n] = \intr \n(\br) \, \epsilon_{\C}\unif[n] \, \tilde G(\zeta, s, \alpha, q) \, ,
\end{equation}
where $\epsilon_{\C}\unif[n]$ is a suitable parameterization for the correlation energy per electron of the uniform gas that has logarithmic divergence in the high-density limit (e.g. PW92~\cite{perdew1992accurate} or VWN~\cite{vosko1980accurate}) and $\tilde G(\zeta, s, \alpha, q)$ depends only on dimensionless variables.  In general, the conventional partitioning we use may differ from the developer's intentions or rationalizations. For instance, approximations that consider a portion of exact exchange to be part of the correlation energy will not be consistent with our conventional partitioning, since such energy contributions will cancel out in Eq.~\eqref{eq:conventional partitioning}.}

{
Throughout, conventional partitioning is utilized whenever the exchange and correlation partitioning is not available in the Libxc library~\cite{LSOM18}. In our analysis, we do not consider range-separated hybrid functionals, double-hybrid functionals, or functionals that contain non-local correlation. Therefore, the correlation energy functional is always expressed in the standard semilocal form considered in this work.
}

\section{Lieb-Oxford bound for hybrid functionals}
\label{app:5}

{
A global hybrid functional satisfies the XC energy LO bound when
\begin{equation}
    \tilde{E}\XC[n] \geq C\LO \, E\X\LDA[n] - \tilde{a} E\X[n] \, .
\label{eq: LO bound hybrids}
\end{equation}
The exchange energy alone follows a tighter bound
\begin{equation}
    E\X[n] \geq C\LO^{\x} \, E\X\LDA[n] \, ,
\end{equation}
with coefficient $C\LO^{\x} < C\LO$. 
Therefore, we can ensure the LO bound with a semilocal functional satisfying
\begin{equation}
\begin{aligned}
\tilde{E}\XC[n] &\geq (C\LO - \tilde{a} \, C\LO^{\x}) \, E\X\LDA[n] \\
&\geq C\LO \, E\X\LDA[n] - \tilde{a} E\X[n] \, . 
\end{aligned}
\label{eq:derive hybrid lo bound}
\end{equation}
Thus, it is sufficient (but not necessary) that $\tilde{E}\XC[n]$ satisfy a LO-like bound with coefficient $C\LO - \tilde{a} \, C\LO^{\x}$. A larger coefficient could result in a violation of the XC energy LO bound in Eq.~\eqref{eq: LO bound hybrids}. To see this, let $\Delta > 0$ and let $n\LO^{\x}$ be a density such that $E\X[n\LO^{\x}] = C\LO \, E\X\LDA[n\LO^{\x}]$, then $(C\LO - \tilde{a} \, C\LO^{\x} + \Delta)E\X\LDA[n\LO^{\x}] < C\LO \, E\X\LDA[n\LO^{\x}] - \tilde{a} E\X[n\LO^{\x}]$. A smaller coefficient would also ensure Eq.~\eqref{eq: LO bound hybrids}, but it would be over-restrictive.}

The corresponding local condition is straightforward
\begin{equation}
    \tilde{F}\XC \leq C\LO - \tilde{a} \, C\LO^{\x} \, .
\label{eq:fxc lob hyb}
\end{equation} 
The corresponding local condition for the LO bound involving $\tilde U\XC\hyb[n]$ is found by using the conventional partitioning in Eq.~\eqref{eq:eps_c conventional partitioning} and applying Eq.~\eqref{eq: Tc def} to yield
\begin{equation}
    \tilde{F}\XC + r_s \frac{\partial \tilde{F}\c}{\partial r_s} \leq C\LO - \tilde{a} \, C\LO^{\x} \, .
\end{equation} 
{In practice, $C\LO$ and $C\LO^{\x}$ are not known precisely and need to be approximated with proven (but not optimal) bounds. 
To give the most benefit of the doubt when assessing approximations, we use $C\LO^{\x} = 1.174$ (the conjectured tight exchange LO coefficient~\cite{perdew2022lieb}) and $C\LO = 2.27$ (which is the same value we use to evaluate non-hybrids). 
}

\section{Local condition for the conjecture: $T\C \leq -E\C$ }
\label{app:6}

While unproven, it has been conjectured~\cite{levy1985hellmann,crisostomo2022seven,frydel2000} that 
\begin{equation}
    T\c[n] \leq -E\c[n] \, \quad \text{(conjecture)} .
\label{eq: tc and uc inequality}
\end{equation}
One can employ the definition in Eq.~\eqref{eq: Tc def} and arrive at the following local condition
\begin{equation}
    \frac{\partial \tilde{F}\c}{\partial r_s} \leq \frac{\tilde{F}\c}{r_s} \, \quad \text{(conjecture)} .
\label{eq:conjecture local}
\end{equation}
In Appendix~\ref{sec: atomic systems} and Supplemental S1 we explore the satisfaction of Eqs.~\eqref{eq: tc and uc inequality} and ~\eqref{eq:conjecture local}, respectively, in approximate functionals. 

\section{Relation to adiabatic connection}
\label{sec:Relation to adiabatic connection}
\label{app:7}

Uniform coordinate scaling is closely related to the adiabatic connection in DFT, which has long been an illuminating concept for rationalizing and improving density functional approximations~\cite{PEB96}. The formalism developed has also revealed many useful exact conditions. 

In the adiabatic connection formalism~\cite{HJ74, LP75, GL76}, we insert a variable coupling constant $\lambda \geq 0$ for Coulomb-interacting electrons with 
\begin{equation}
    F^{\lambda}[\n] = \min_{\Psi \rightarrow \n} \bra{\Psi} \hat{T} + \lambda \, \hat{V}\ee \ket{\Psi} \, ,
\label{eq: universal functional lambda}
\end{equation}
where $\hat{T}$ is the usual total kinetic energy operator, $\hat{V}\ee$ is the two-body electron-electron repulsion operator, and the minimization is over all antisymmetric wavefunctions that yield the density $n$. For $\lambda = 1$ and ground-state density $n(\br)$, we have our real physical system. Taking $\lambda = 0$, we have the KS system, and $F^{\lambda = 0}[n] = T_s[n]$, where $T_s[n]$ is the kinetic energy of the non-interacting KS system with ground-state density $n(\br)$. In all cases $\lambda \geq 0$, the ground-state density remains fixed to that of the physical system $n(\br)$. 

The adiabatic connection is directly coupled to uniform coordinate scaling by
\begin{equation}
    F^{\lambda}[n] = \lambda^2 F[n_{1/\lambda}] \, .
\end{equation}
This relation also extends to any other energy functional component, e.g., $E\C^{\lambda}[n] = \lambda^2 E\C[n_{1/\lambda}]$. Therefore,  exact conditions written in terms of uniform coordinate scaling can be recast in terms of adiabatic connection quantities.

Eq.~\eqref{eq: correlation uniform scaling inequality} is rewritten as 
\begin{equation}
\begin{aligned}
    E\C^{\lambda}[n] \geq \lambda \, E\C[n]  \qquad (\lambda < 1) \, ,\\
    E\C^{\lambda}[n] \leq \lambda \, E\C[n]  \qquad (\lambda > 1) \, .
\end{aligned}
\label{eq: correlation lambda inequality}
\end{equation}
Using
\begin{equation}
    \frac{\partial E\C[n_{1/\lambda}]}{\partial (1/\lambda)} = - \frac{\partial E\C^{\lambda}[n]}{\partial \lambda} + 2 \frac{E\C^{\lambda}[n]}{\lambda}
\end{equation}
we recast the exact condition in Eq.~\eqref{eq:Tc upperbound exact cond} as
\begin{equation}
    \frac{\partial E\C^{\lambda}[n]}{\partial \lambda} \geq \bigg( -\frac{\partial E\C^{\lambda}[n]}{\partial \lambda} + 2 \frac{E\C^{\lambda}[n]}{\lambda} \bigg) \Bigr|_{\lambda \to \infty} \, ,
\end{equation}
and applying L'Hospital's rule we obtain
\begin{equation}
    \frac{\partial E\C^{\lambda}[n]}{\partial \lambda} \geq \frac{\partial E\C^{\lambda}[n]}{\partial \lambda} \Bigr|_{\lambda \to \infty} \, .
\end{equation}
This equation can also be rewritten as a simple statement in the \textit{strictly correlated electron} (SCE)~\cite{seidl2007strictly} limit ($\lambda \to \infty$),
\begin{equation}
    \bra{\Psi^{\lambda}[n]} \hat{V}\ee \ket{\Psi^{\lambda}[n]} \geq \bra{\Psi^{\lambda \to \infty}[n]} \hat{V}\ee \ket{\Psi^{\lambda \to \infty}[n]} \, .
\label{eq:sce vee inequality}
\end{equation}
In the SCE limit, the kinetic energy component is subleading and $\Psi^{\lambda \to \infty}[n]$ minimizes $\hat{V}\ee$, thus Eq.~\eqref{eq:sce vee inequality} is clear.

\section{Analytical derivation: PBE satisfies the correlation uniform scaling inequality exact condition}
\label{app:8}

The PBE correlation energy functional has the form~\cite{PBE96}
\begin{equation}
    E^{\text{PBE}}\C[n] = \intr n(\br) \, \big( \epsilon^{\text{PW92}}\C(r_s, \zeta) +  H(r_s, \zeta, t) \big) \, ,
\end{equation}
where $\epsilon^{\text{PW92}}\C$ is the PW92~\cite{perdew1992accurate} parameterized correlation energy per electron of the uniform gas and $H(r_s, \zeta, t)$ is defined in Ref.~\cite{PBE96} along with the dimensionless gradient $t$. For simplicity, we set all positive constants in $H$ to unity, as the final conclusion will not depend on their specific values. With this,
\begin{equation}
    H = \ln \bigg[1 + t^2 \bigg[ \frac{1 + At^2}{1 + At^2 + A^2 t^4} \bigg] \bigg] \geq 0 \, , 
\end{equation}
where
\begin{equation}
    A = [\exp(-\epsilon_c^{\text{PW92}}) - 1]^{-1} \geq 0 \, .
\end{equation}
In the following, we show that PBE correlation satisfies Eq.~\eqref{eq:Fc lower bd cond}, and thus the correlation uniform scaling inequality exact condition in Eq.~\eqref{eq: correlation uniform scaling inequality}. Since PW92 already satisfies Eq.~\eqref{eq:Fc lower bd cond}, proving the following condition,
\begin{equation}
    \frac{\partial}{\partial r_s} \bigg( \frac{H(r_s, \zeta, t)}{\epsilon_{\x}^{\text{unif}}(r_s)}\bigg) \geq 0 \, ,
\label{eq:}
\end{equation}
is sufficient to ensure that PBE satisfies the exact condition.

To show this, we start with
\begin{equation}
    \frac{\partial}{\partial r_s} \bigg( \frac{H(r_s, \zeta, t)}{\epsilon_{\x}^{\text{unif}}(r_s)}\bigg) = \frac{(\frac{\partial}{\partial r_s} H)}{\epsilon_{\x}^{\text{unif}}} + \frac{(\frac{\partial}{\partial r_s} \epsilon_{\x}^{\text{unif}}) H}{(\epsilon_{\x}^{ \text{unif}})^2} \, .
\label{eq: pbe quotient rule}
\end{equation}
Since $\epsilon_{\x}^{\text{unif}} = -(3/4) (3/2 \pi)^{2/3} r_s^{-1}$, the second term in Eq.~\eqref{eq: pbe quotient rule} is positive. Next we evaluate the first term, 
\begin{equation}
    \frac{\partial H}{\partial r_s} = \frac{1}{1+x} \times \frac{\partial x}{\partial A} \times \frac{\partial A}{\partial \epsilon_c^{\text{PW92}}} \times \frac{\partial \epsilon_c^{\text{PW92}}}{\partial r_s} \, ,
\end{equation}
where 
\begin{equation}
    x = t^2 \bigg[ \frac{1 + A t^2}{1 + A t^2 + A^2 t^4} \bigg] \geq 0 \, .
\end{equation}
The intermediate derivatives are derived analytically:
\begin{equation}
    \frac{\partial x}{\partial A} = \frac{- t^4 A ( t^2 A + 2)}{(t^4 A^2 + t^2 A + 1)^2} \leq 0 \, ,
\end{equation}
\begin{equation}
    \frac{\partial A}{\partial \epsilon_c^{\text{PW92}}} = \frac{\exp(- \epsilon_c^{\text{PW92}})}{(\exp(- \epsilon_c^{\text{PW92}}) - 1)^2} \geq 0 \, ,
\end{equation}
and
\begin{equation}
    \frac{\partial \epsilon_{\c}^{\text{PW92}}}{\partial r_s} \geq 0 \, .
\end{equation}
Therefore, 
\begin{equation}
    \frac{\partial H}{\partial r_s} \leq 0 \, , 
\end{equation}
and the first term in Eq.~\eqref{eq: pbe quotient rule} is also positive and thus the exact condition is satisfied.

\section{Readjusting parameters in the SCAN functional}
\label{app:9}

While the local condition in Eq.~\eqref{eq:Fc lower bd cond} is satisfied in the published SCAN functional, reasonable adjustments of the parameters designated for the fitting of the ``appropriate norms'' can result in violations of the local condition. Specifically, we adjust the $b_{1c} = 0.02858$ parameter in SCAN, which was fit to match the correlation energy of the $Z \to \infty$ limit of two-electron ions (one of the five appropriate norms in SCAN, see the supplementary material in Ref.~\cite{scan_functional}). By evaluating SCAN analytically, we find that using $b_{1c} > 0.2$ results in violations of Eq.~\eqref{eq:Fc lower bd cond}. Under this single parameter modification, the other exact conditions that SCAN satisfies are still satisfied by construction. Therefore, by virtue of fitting to various appropriate norms, the SCAN functional satisfies more exact conditions than were explicitly enforced.

\section{exact conditions on atomic system densities}
\label{app:10}
\label{sec: atomic systems}

\begin{figure}[h]
\includegraphics[width=0.4\textwidth]{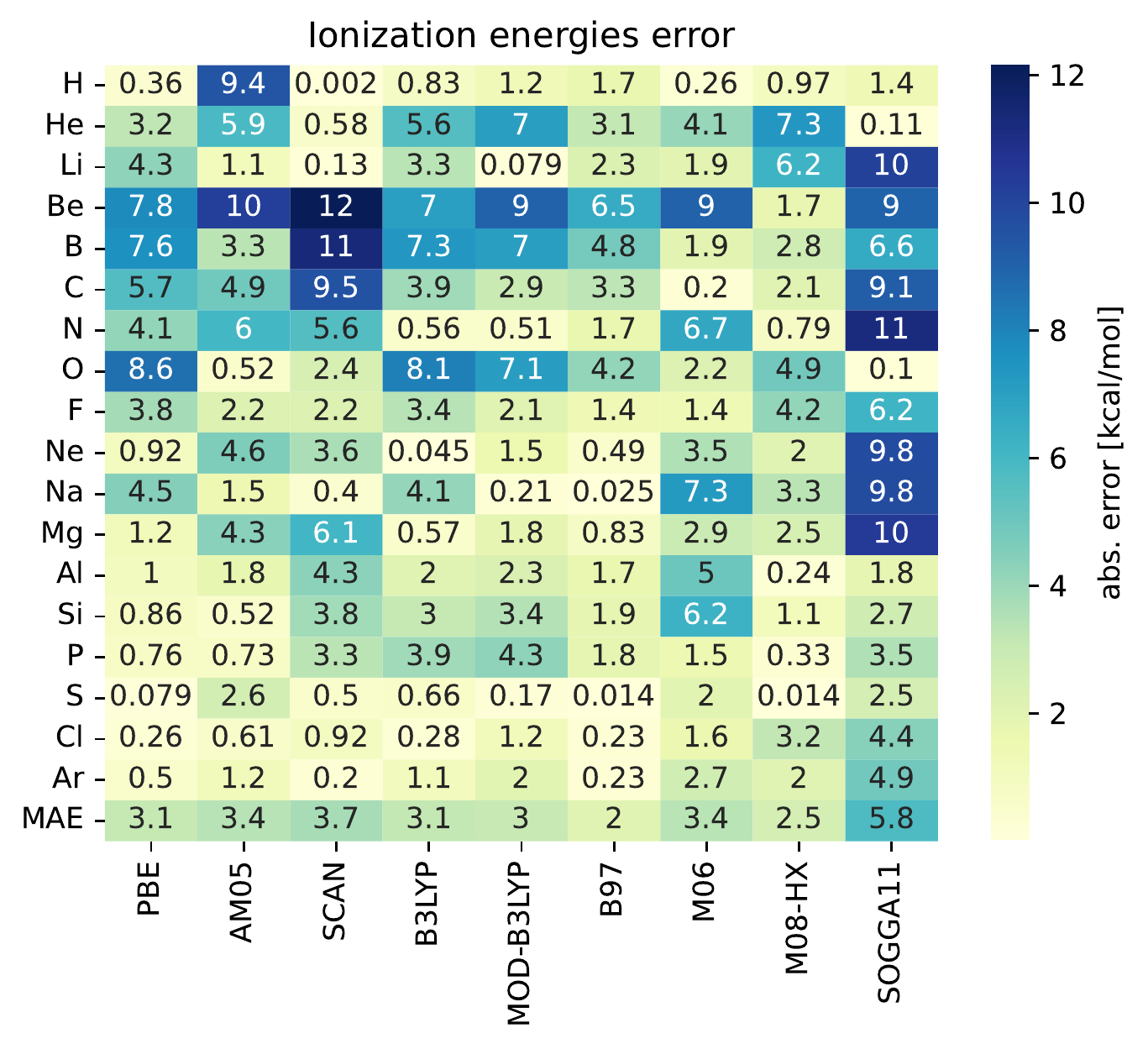}
\centering
\caption{Ionization energy errors for atomic systems H-Ar. All approximations are evaluated using non-self-consistent $\Delta$SCF~\cite{jones1989density} with HF neutral and +1 cationic densities and orbitals.}
\label{fig: IE error heatmap}
\end{figure}

{
In this assessment, we first calculate Hartree-Fock (HF) densities and orbitals for neutral atoms H-Ar and their cations. The fixed HF densities and orbitals are then used to evaluate the energies (non-self-consistently) from different DFT approximations. 
HF densities are used because they provide high quality densities and an equal footing across different approximations. 
We also performed separate calculations using self-consistent densities and observed marginal differences.
The absolute errors from the experimental ionization energies are provided in Fig.~\ref{fig: IE error heatmap}.
}
{In addition to established approximations, we also test a very simple modified B3LYP (MOD-B3LYP) which satisfies the correlation exact conditions discussed for \textit{any} density,
\begin{equation}
    \epsilon^{\text{MOD-B3LYP}}\C(r_s, \zeta, s) = \Theta(s - 1.82) \, \epsilon^{\text{B3LYP}}\C(r_s, \zeta, s) \, ,
\end{equation}
where $\epsilon\c^{\text{B3LYP}} = 0.405 \epsilon\c^{\text{LYP}} +  0.095 \epsilon\c^{\text{VWN5}}$~\cite{stephens1994ab}.The step function {eliminates} the local condition violations found in Table~1 of the main text. 
As argued in the main text, energy contributions from such large $s$ values are less relevant in Coulombic systems, and indeed in Fig.~\ref{fig: IE error heatmap} we see that B3LYP and MOD-B3LYP have MAEs that differ only by $0.1$ kcal/mol. Our modified functional is constructed for demonstration purposes only.}

\begin{figure}[h]
\includegraphics[width=0.4\textwidth]{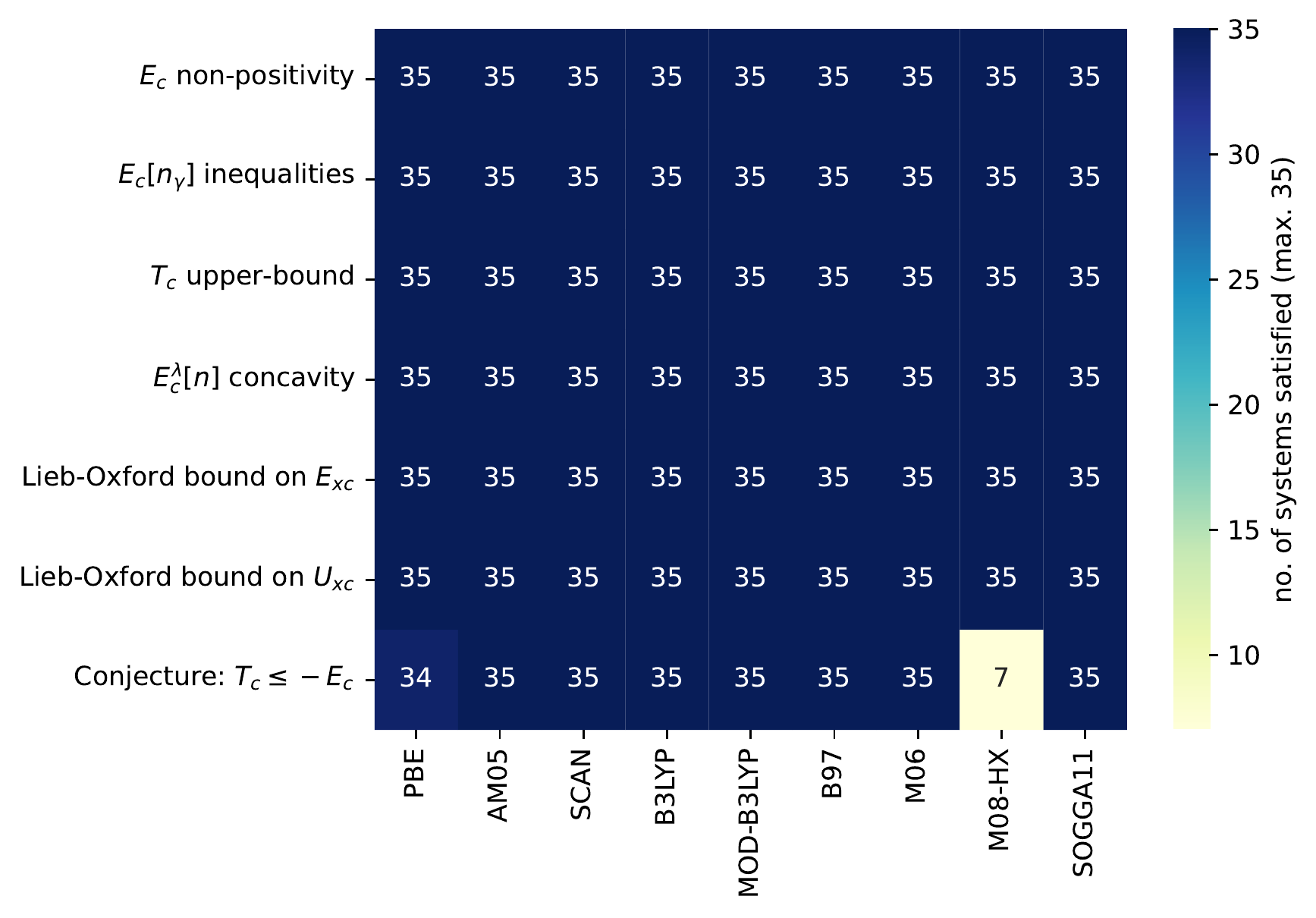}
\centering
\caption{{The number of HF atomic densities tested which satisfy a given exact condition. A total of $35$ atomic systems are tested: H-Ar and their cations.}}
\label{fig: exact cond heatmap}
\end{figure}

{For each HF density (a total of 35 atomic systems), we scale the density $n\g$ with $\gamma \in [0.01, 2]$ (50 evenly spaced values) and evaluate whether an exact condition is satisfied. Indeed, in Fig.~\ref{fig: exact cond heatmap} we see that the exact conditions tested are all satisfied within our set of functional approximations and atomic systems. We also test the conjecture $T_c \leq -E_c$ and find instances of violation for PBE and M08-HX.}

\clearpage

\bibliographystyle{apsrev4-2}
\bibliography{Master,ryan_master}

\label{page:end}
\end{document}

\sf
\coloredtitle{\TitleOfPaper}

%Put the names of authors here:
\coloredauthor{Ryan Pederson}
\email{pedersor@uci.edu}
\affiliation{Department of Physics and Astronomy, University of California, Irvine, CA 92697, USA}

\coloredauthor{Kieron Burke}
\email{kieron@uci.edu}
\affiliation{Department of Chemistry, University of California, Irvine, CA 92697, USA}
\affiliation{Department of Physics and Astronomy, University of California, Irvine, CA 92697, USA}

\date{\today}

\maketitle

\tableofcontents

\sec{assessment of local conditions across available approximations }
\label{sec: local condition tables}

We utilize an exhaustive grid search to determine whether local conditions are satisfied for a given approximation. \edit{For LDAs, we consider $10000$ evenly spaced values of $r_s \in [0.0001, 5]$ and $100$ evenly spaced values of $\zeta \in [0,1]$.} For GGAs, we consider $10000$ evenly spaced values of $r_s \in [0.0001, 5]$, $500$ evenly spaced values of $s \in [0, 5]$, and $100$ evenly spaced values of $\zeta \in [0,1]$. For MGGAs, we consider $5000$ evenly spaced values of $r_s \in [0.0001, 5]$, $100$ evenly spaced values of $s \in [0, 5]$, $20$ evenly spaced values of $\zeta \in [0,1]$, and $100$ evenly spaced values of $\alpha \in [0, 5]$ or $q \in [-10, 10]$. 
The number of values checked per variable is less in MGGAs to alleviate computational effort due to the combinatorial nature of the exhaustive search. 

In determining whether local conditions are satisfied, a reasonable tolerance threshold of at most $\pm 0.001$ is employed to approximately account for numerical errors arising from the numerical precision used and the finite difference method used to calculate numerical derivatives (further details can be found in our public code~\cite{dft_exconditions}). However, the numerical errors introduced are not guaranteed to be within the tolerances used.

In the following tables below, we report the fraction of local condition violations found in our exhaustive search. That is, we divide the number of violations found by the total number of configurations considered in the extensive grid search parameter space. If we find $0$ such violations, then we conclude that the corresponding exact condition is satisfied for any reasonable density. \edit{We do not assess the LO bound condition on approximations that are for correlation only.}

\begin{table*}
\caption{LDA functionals: numerical assessment of corresponding local conditions.}
\begin{tabular}{|l|c|c|c|c|c|c|c|}
\toprule
 & \makecell[c]{$E\C[n]$ \\ non-positivity} & \makecell[c]{$E\C[n\g]$ uniform \\ scaling inequality} & \makecell[c]{$T\C[n]$ \\ upper bound} & \makecell[c]{concavity of \\ $\lambda^2 E\c[n_{1/\lambda}]$} & \makecell[c]{LO extension \\ to $E\xc$} & \makecell[c]{LO} & \makecell[c]{conjecture: \\ $T_c \leq -E_c$} \\
\midrule
BR78~\cite{Brual1978_1177} & 0 & 0 & 0 & 0 & --- & --- & 0 \\
CHACHIYO-MOD~\cite{Chachiyo2020_112669,Chachiyo2016_021101} & 0 & 0 & 0 & 0 & --- & --- & 0 \\
CHACHIYO~\cite{Chachiyo2016_021101} & 0 & 0 & 0 & 0 & --- & --- & 0 \\
GDSMFB~\cite{Groth2017_135001} & 1e-04 & 2e-05 & 0 & 0 & 0 & 0 & 6e-04 \\
GK72~\cite{Gordon1972_3122} & 0 & 1e-04 & 0.999 & 1e-04 & --- & --- & 0 \\
GL~\cite{Gunnarsson1976_4274} & 0 & 0 & 0 & 0 & --- & --- & 0 \\
GOMBAS~\cite{Gombas1965_137} & 0 & 0 & 0 & 0 & --- & --- & 0 \\
HL~\cite{Hedin1971_2064} & 0 & 0 & 0 & 0 & --- & --- & 0 \\
KARASIEV-MOD~\cite{Chachiyo2020_112669,Karasiev2016_157101} & 0 & 0 & 0 & 0 & --- & --- & 0 \\
KARASIEV~\cite{Karasiev2016_157101} & 0 & 0 & 0 & 0 & --- & --- & 0 \\
KSDT~\cite{Karasiev2014_076403} & 0 & 0 & 0 & 0 & 0 & 0 & 0 \\
LDA0~\cite{Rinke2012_126404} & 0 & 0 & 0 & 0 & 0 & 0 & 0 \\
LP-A~\cite{Lee1990_193} & 0 & 0 & 0 & 0 & 0 & 0 & 0 \\
LP-B~\cite{Lee1990_193} & 0 & 0 & 0.989 & 0.935 & 0 & 0 & 0 \\
LP96~\cite{Liu1996_2211,Liu2000_29} & 0.515 & 0.726 & 0 & 0.845 & --- & --- & 0 \\
MCWEENY~\cite{McWeeny1976_3,Brual1978_1177} & 0 & 0 & 0 & 0 & --- & --- & 0 \\
ML1~\cite{Proynov1994_7874,Proynov1998_12616} & 0 & 9e-05 & 0 & 0.004 & --- & --- & 6e-05 \\
ML2~\cite{Proynov1994_7874,Proynov1998_12616} & 0 & 9e-05 & 0 & 0.004 & --- & --- & 8e-06 \\
OB-PW~\cite{Ortiz1994_1391,Ortiz1994_1391_err,Perdew1992_13244_mod} & 0 & 0 & 0 & 0 & --- & --- & 0 \\
OB-PZ~\cite{Ortiz1994_1391,Ortiz1994_1391_err} & 0 & 0 & 0 & 1e-04 & --- & --- & 0 \\
OW-LYP~\cite{Stewart1995_4337} & 0 & 0 & 0 & 0 & --- & --- & 0 \\
OW~\cite{Stewart1995_4337} & 0 & 0 & 0 & 0 & --- & --- & 0 \\
PK09~\cite{Proynov2009_014103,Proynov2017_059904} & 0 & 0 & 0 & 4e-04 & --- & --- & 0 \\
PMGB06~\cite{Paziani2006_155111} & 0 & 0 & 0 & 0 & --- & --- & 0.970 \\
PW-MOD~\cite{Perdew1992_13244_mod} & 0 & 0 & 0 & 0 & --- & --- & 0 \\
PW-RPA~\cite{Perdew1992_13244} & 0 & 0 & 0 & 0 & --- & --- & 0 \\
PW~\cite{Perdew1992_13244} & 0 & 0 & 0 & 0 & --- & --- & 0 \\
PZ-MOD~\cite{Perdew1981_5048_mod} & 0 & 0 & 0 & 0 & --- & --- & 0 \\
PZ~\cite{Perdew1981_5048} & 0 & 0 & 0 & 1e-04 & --- & --- & 1e-04 \\
RC04~\cite{Ragot2004_7671} & 0 & 0 & 0 & 0 & --- & --- & 0 \\
RPA~\cite{GellMann1957_364} & 0 & 0.336 & 0.999 & 0.602 & --- & --- & 0 \\
RPW92~\cite{Ruggeri2018_161105} & 0 & 0 & 0 & 0 & --- & --- & 0 \\
TETER93~\cite{Goedecker1996_1703} & 0 & 0 & 0 & 0 & 0 & 0 & 0 \\
UPW92~\cite{Ruggeri2018_161105} & 0 & 0 & 0 & 0 & --- & --- & 0 \\
VBH~\cite{vonBarth1972_1629} & 0 & 0 & 0 & 0 & --- & --- & 0 \\
VWN-1~\cite{Vosko1980_1200} & 0 & 0 & 0 & 0 & --- & --- & 0 \\
VWN-2~\cite{Vosko1980_1200} & 0 & 0 & 0 & 0 & --- & --- & 0 \\
VWN-3~\cite{Vosko1980_1200} & 0 & 0 & 0 & 0 & --- & --- & 0 \\
VWN-4~\cite{Vosko1980_1200} & 0 & 0 & 0 & 0 & --- & --- & 0 \\
VWN-RPA~\cite{Vosko1980_1200} & 0 & 0 & 0 & 0 & --- & --- & 0 \\
VWN~\cite{Vosko1980_1200} & 0 & 0 & 0 & 0 & --- & --- & 0 \\
W20~\cite{Xie2021_045130} & 0 & 0 & 0 & 0 & --- & --- & 0 \\
WIGNER~\cite{Wigner1938_678,Stewart1995_4337} & 0 & 0 & 0 & 0 & --- & --- & 0 \\
XALPHA~\cite{Slater1951_385} & 0 & 0 & 0 & 0 & --- & --- & 0 \\
ZLP~\cite{Zhao1993_918} & 0 & 0 & 0.999 & 0.915 & 0 & 0 & 0 \\
\bottomrule
\end{tabular}
\end{table*}

\begin{table*}
\caption{GGA functionals: numerical assessment of corresponding local conditions.}
\begin{tabular}{|l|c|c|c|c|c|c|c|}
\toprule
 & \makecell[c]{$E\C[n]$ \\ non-positivity} & \makecell[c]{$E\C[n\g]$ uniform \\ scaling inequality} & \makecell[c]{$T\C[n]$ \\ upper bound} & \makecell[c]{concavity of \\ $\lambda^2 E\c[n_{1/\lambda}]$} & \makecell[c]{LO extension \\ to $E\xc$} & \makecell[c]{LO} & \makecell[c]{conjecture: \\ $T_c \leq -E_c$} \\
\midrule
ACGGAP~\cite{Cancio2018_084116,Burke2014_4834} & 0 & 0 & 0 & 0 & --- & --- & 0.414 \\
ACGGA~\cite{Cancio2018_084116,Burke2014_4834} & 0 & 0 & 0 & 0 & --- & --- & 0 \\
AM05~\cite{Armiento2005_085108,Mattsson2008_084714} & 0 & 0 & 0 & 0 & 0 & 0 & 0 \\
APBE~\cite{Constantin2011_186406} & 0 & 0 & 0 & 0 & 0 & 0 & 0.004 \\
B97-D~\cite{Grimme2006_1787} & 0.632 & 0.503 & 0.559 & 0.608 & 0.268 & 0.183 & 0.633 \\
B97-GGA1~\cite{Cohen2000_160} & 0.636 & 0.514 & 0.557 & 0.612 & 0.390 & 0.317 & 0.639 \\
BEEFVDW~\cite{Wellendorff2012_235149} & 0 & 0 & 0 & 0 & 0.003 & 0.013 & 0 \\
BMK~\cite{Boese2004_3405} & 0.627 & 0.304 & 0.648 & 0.621 & --- & --- & 0.616 \\
CCDF~\cite{Margraf2019_244116} & 0 & 0 & 0 & 0 & --- & --- & 0 \\
CHACHIYO~\cite{Chachiyo2020_112669} & 0 & 0 & 0.044 & 0 & 0.217 & 0.217 & 0.010 \\
CS1~\cite{Handy2002_5411,Proynov2006_436} & 0.604 & 0.204 & 0.528 & 0.530 & --- & --- & 0.601 \\
EDF1~\cite{Adamson1998_6} & 0.605 & 0.245 & 0.002 & 0.231 & 0.162 & 0.203 & 0.527 \\
FT97~\cite{Filatov1997_603,Filatov1997_847} & 0 & 0 & 1e-05 & 0.003 & --- & --- & 0 \\
GAM~\cite{Yu2015_12146} & 0.598 & 0.459 & 0.560 & 0.578 & 0.145 & 0.083 & 0.596 \\
GAPC~\cite{Fabiano2014_2016} & 0.004 & 0.011 & 2e-04 & 0.005 & --- & --- & 0.015 \\
GAPLOC~\cite{Fabiano2014_2016} & 4e-04 & 2e-04 & 2e-04 & 0.005 & --- & --- & 0.033 \\
HCTH-120~\cite{Boese2000_1670} & 0.495 & 0.310 & 0.327 & 0.450 & 0.065 & 0.061 & 0.507 \\
HCTH-147~\cite{Boese2000_1670} & 0.467 & 0.290 & 0.298 & 0.422 & 0.113 & 0.093 & 0.478 \\
HCTH-407P~\cite{Boese2003_5965} & 0.536 & 0.445 & 0.428 & 0.508 & 0.105 & 0.075 & 0.543 \\
HCTH-407~\cite{Boese2001_5497} & 0.481 & 0.382 & 0.365 & 0.450 & 0.112 & 0.079 & 0.489 \\
HCTH-93~\cite{Hamprecht1998_6264} & 0.435 & 0.196 & 0.258 & 0.386 & 0.266 & 0.237 & 0.446 \\
HCTH-A~\cite{Hamprecht1998_6264} & 0.493 & 0.289 & 0.355 & 0.454 & 0 & 0 & 0.501 \\
HCTH-P14~\cite{Menconi2001_3958} & 0 & 0 & 0 & 0 & 0 & 0.029 & 0 \\
HCTH-P76~\cite{Menconi2001_3958} & 0.986 & 0.929 & 0.999 & 0.991 & 0.011 & 0.005 & 0.978 \\
HLE16~\cite{Verma2017_380} & 0.481 & 0.305 & 0.364 & 0.447 & 0.477 & 0.473 & 0.487 \\
HYB-TAU-HCTH~\cite{Boese2002_9559} & 0.615 & 0.439 & 0.520 & 0.585 & --- & --- & 0.620 \\
KT1~\cite{Keal2003_3015} & 0.791 & 0.434 & 0.096 & 0.169 & 0.153 & 0.076 & 0.664 \\
KT2~\cite{Keal2003_3015} & 0.832 & 0.477 & 0.106 & 0.177 & 0.156 & 0.077 & 0.686 \\
KT3~\cite{Keal2004_5654} & 0.862 & 0.461 & 0.120 & 0.192 & 0.164 & 0.077 & 0.678 \\
LM~\cite{Langreth1981_446,Hu1985_391} & 0 & 0.119 & 0.464 & 0.384 & --- & --- & 0 \\
LYPR~\cite{Ai2021_1207} & 0.320 & 0.113 & 0.801 & 0.590 & --- & --- & 0.438 \\
LYP~\cite{Lee1988_785,Miehlich1989_200} & 0.576 & 0.218 & 0.003 & 0.203 & --- & --- & 0.511 \\
MGGAC~\cite{Patra2019_155140} & 0 & 0 & 0 & 0 & --- & --- & 0.007 \\
MOHLYP2~\cite{Zheng2009_808} & 0.576 & 0.174 & 0.002 & 0.193 & 0.340 & 0.337 & 0.509 \\
MOHLYP~\cite{Schultz2005_11127} & 0.243 & 0.092 & 0 & 0.092 & 0.048 & 0.096 & 0.328 \\
MPWLYP1W~\cite{Dahlke2005_15677} & 0.500 & 0.190 & 4e-07 & 0.168 & 0.003 & 0.004 & 0.474 \\
N12~\cite{Peverati2012_2310} & 0 & 0 & 0 & 0 & 0.150 & 0.170 & 0 \\
NCAP~\cite{Carmona2019_303} & 0.455 & 0.299 & 0.029 & 0.277 & 0.231 & 0.207 & 0.403 \\
OBLYP-D~\cite{Goerigk2010_107} & 0.595 & 0.246 & 0.002 & 0.243 & 0.019 & 0.020 & 0.505 \\
OP-B88~\cite{Tsuneda1999_10664} & 0 & 2e-04 & 6e-04 & 0.002 & --- & --- & 7e-04 \\
OP-G96~\cite{Tsuneda1999_10664,Tsuneda1999_5656} & 0 & 2e-04 & 6e-04 & 0.002 & --- & --- & 7e-04 \\
OP-PBE~\cite{Tsuneda1999_10664,Tsuneda1999_5656} & 0 & 2e-04 & 6e-04 & 0.002 & --- & --- & 7e-04 \\
OP-PW91~\cite{Tsuneda1999_10664,Tsuneda1999_5656} & 0 & 0.001 & 0.001 & 0.002 & --- & --- & 0.001 \\
OP-XALPHA~\cite{Tsuneda1999_10664,Tsuneda1999_5656} & 0 & 2e-04 & 6e-04 & 0.002 & --- & --- & 7e-04 \\
OPBE-D~\cite{Goerigk2010_107} & 0 & 0 & 0 & 0 & 0.009 & 0.009 & 0.006 \\
OPTC~\cite{Cohen2001_607} & 3e-06 & 0 & 0 & 0 & --- & --- & 0 \\
OPWLYP-D~\cite{Goerigk2010_107} & 0.596 & 0.247 & 7e-04 & 0.240 & 0.019 & 0.022 & 0.507 \\
P86-FT~\cite{Perdew1986_8822} & 0.454 & 0.298 & 0.027 & 0.276 & --- & --- & 0.402 \\
P86VWN-FT~\cite{Perdew1986_8822} & 0.447 & 0.297 & 0.026 & 0.275 & --- & --- & 0.389 \\
P86VWN~\cite{Perdew1986_8822} & 0.447 & 0.297 & 0.026 & 0.275 & --- & --- & 0.389 \\
P86~\cite{Perdew1986_8822} & 0.454 & 0.298 & 0.027 & 0.276 & --- & --- & 0.403 \\
PBE-JRGX~\cite{Pedroza2009_201106} & 0 & 0 & 0 & 0 & --- & --- & 0.006 \\
PBE-MOL~\cite{delCampo2012_104108} & 0 & 0 & 0 & 0 & 0 & 0 & 0.004 \\
PBE-SOL~\cite{Perdew2008_136406} & 0 & 0 & 0 & 0 & 0 & 0 & 0.006 \\
PBE-VWN~\cite{Kraisler2010_042516,Perdew1996_3865,Perdew1997_1396} & 0 & 0 & 0 & 0 & --- & --- & 6e-04 \\
PBE1W~\cite{Dahlke2005_15677} & 0 & 0 & 0 & 0 & 0 & 0 & 0 \\
PBEFE~\cite{Perez2015_3844} & 0 & 0 & 0 & 0 & 0 & 0 & 0.006 \\
PBEINT~\cite{Fabiano2010_113104} & 0 & 0 & 0 & 0 & 0 & 0 & 0.005 \\
PBELOC~\cite{Constantin2012_035130} & 0 & 0 & 0 & 0.003 & --- & --- & 0.271 \\
PBELYP1W~\cite{Dahlke2005_15677} & 0.414 & 0.157 & 0 & 0.141 & 0 & 0 & 0.427 \\
PBE~\cite{Perdew1996_3865,Perdew1997_1396} & 0 & 0 & 0 & 0 & 0 & 0 & 0.005 \\
\bottomrule
\end{tabular}
\end{table*}

\begin{table*}
\caption{GGA functionals: numerical assessment of corresponding local conditions.}
\begin{tabular}{|l|c|c|c|c|c|c|c|}
\toprule
 & \makecell[c]{$E\C[n]$ \\ non-positivity} & \makecell[c]{$E\C[n\g]$ uniform \\ scaling inequality} & \makecell[c]{$T\C[n]$ \\ upper bound} & \makecell[c]{concavity of \\ $\lambda^2 E\c[n_{1/\lambda}]$} & \makecell[c]{LO extension \\ to $E\xc$} & \makecell[c]{LO} & \makecell[c]{conjecture: \\ $T_c \leq -E_c$} \\
\midrule
PW91~\cite{Perdew1991,Perdew1992_6671,Perdew1993_4978} & 3e-06 & 0 & 0 & 0 & 0 & 0 & 0 \\
Q2D~\cite{Chiodo2012_126402} & 0 & 0.041 & 0.012 & 0.032 & 0 & 0 & 0.002 \\
REGTPSS~\cite{Perdew2009_026403} & 0 & 0 & 0 & 0 & --- & --- & 0.406 \\
REVTCA~\cite{Tognetti2008_536} & 0 & 0 & 0.003 & 0.051 & --- & --- & 0.024 \\
RGE2~\cite{Ruzsinszky2009_763} & 0 & 0 & 0 & 0 & 0 & 0 & 0.005 \\
SCAN-E0~\cite{Sun2015_036402} & 0 & 0 & 0 & 0 & --- & --- & 0 \\
SG4~\cite{Constantin2016_045126} & 0 & 0.048 & 0.327 & 0.385 & 8e-04 & 0.007 & 0.050 \\
SOGGA11~\cite{Peverati2011_1991} & 0 & 0.003 & 0.064 & 0.229 & 0 & 1e-04 & 0.002 \\
SPBE~\cite{Swart2009_094103} & 0 & 0 & 0 & 0 & --- & --- & 0 \\
TAU-HCTH~\cite{Boese2002_9559} & 0.595 & 0.491 & 0.492 & 0.567 & --- & --- & 0.603 \\
TCA~\cite{Tognetti2008_034101} & 0 & 0 & 0 & 0 & --- & --- & 0 \\
TH-FCFO~\cite{Tozer1997_183} & 0.223 & 0.772 & 0.233 & 0.237 & 0.787 & 0.228 & 0.226 \\
TH-FCO~\cite{Tozer1997_183} & 0.200 & 0.795 & 0.211 & 0.214 & 0.781 & 0.205 & 0.203 \\
TH-FC~\cite{Tozer1997_183} & 0.988 & 0.009 & 0.994 & 0.996 & 0.822 & 0.990 & 0.989 \\
TH-FL~\cite{Tozer1997_183} & 0 & 1.000 & 0 & 0 & 0.498 & 0 & 0 \\
TH1~\cite{Tozer1998_2545} & 0.215 & 0.780 & 0.224 & 0.225 & 0.295 & 0.220 & 0.218 \\
TH2~\cite{Tozer1998_3162} & 0.061 & 0.935 & 0.070 & 0.070 & 0.323 & 0.065 & 0.063 \\
TH3~\cite{Handy1998_707} & 0.217 & 0.781 & 0.217 & 0.219 & 0.284 & 0.218 & 0.218 \\
TH4~\cite{Handy1998_707} & 0.103 & 0.894 & 0.106 & 0.106 & 0.238 & 0.105 & 0.105 \\
TM-LYP~\cite{Thakkar2009_134109} & 0.575 & 0.209 & 0.122 & 0.173 & --- & --- & 0.565 \\
TM-PBE~\cite{Thakkar2009_134109} & 0 & 0 & 0 & 0 & --- & --- & 0.509 \\
W94~\cite{Wilson1994_337} & 0 & 0 & 0 & 3e-05 & --- & --- & 0 \\
WI0~\cite{Wilson1998_523} & 0.614 & 0.004 & 0.002 & 0.014 & --- & --- & 0.603 \\
WI~\cite{Wilson1998_523} & 0.900 & 0.008 & 0.008 & 0.023 & --- & --- & 0.896 \\
WL~\cite{Wilson1990_12930} & 0.590 & 0.166 & 0.595 & 0.581 & --- & --- & 0.377 \\
XLYP~\cite{Xu2004_2673} & 0.576 & 0.218 & 0.003 & 0.203 & 0.013 & 0.019 & 0.511 \\
XPBE~\cite{Xu2004_4068} & 0 & 0 & 0 & 0 & 0 & 0 & 0 \\
ZPBEINT~\cite{Constantin2011_233103} & 0 & 0.020 & 0.343 & 0.299 & --- & --- & 2e-04 \\
ZPBESOL~\cite{Constantin2011_233103} & 0 & 0.013 & 0.375 & 0.263 & --- & --- & 2e-04 \\
ZVPBEINT~\cite{Constantin2012_194105} & 0 & 0.093 & 0.273 & 0.226 & --- & --- & 0.001 \\
ZVPBELOC~\cite{Fabiano2015_122} & 0 & 3e-04 & 0.199 & 0.112 & --- & --- & 0.075 \\
ZVPBESOL~\cite{Constantin2012_194105} & 0 & 0.080 & 0.299 & 0.212 & --- & --- & 0.001 \\
\bottomrule
\end{tabular}
\end{table*}

\begin{table*}
\caption{MGGA functionals: numerical assessment of corresponding local conditions.}
\begin{tabular}{|l|c|c|c|c|c|c|c|}
\toprule
 & \makecell[c]{$E\C[n]$ \\ non-positivity} & \makecell[c]{$E\C[n\g]$ uniform \\ scaling inequality} & \makecell[c]{$T\C[n]$ \\ upper bound} & \makecell[c]{concavity of \\ $\lambda^2 E\c[n_{1/\lambda}]$} & \makecell[c]{LO extension \\ to $E\xc$} & \makecell[c]{LO} & \makecell[c]{conjecture: \\ $T_c \leq -E_c$} \\
\midrule
B88~\cite{Becke1988_1053} & 0 & 0 & 0 & 0 & --- & --- & 0 \\
B94~\cite{Becke1994_625} & 0 & 0 & 0 & 0 & --- & --- & 0 \\
BC95~\cite{Becke1996_1040} & 0 & 0 & 0 & 0 & --- & --- & 0 \\
CC06~\cite{Cancio2006_081202} & 0 & 0 & 0 & 0 & 0 & 0 & 0 \\
CS~\cite{Colle1975_329,Lee1988_785} & 0.352 & 0.166 & 0.252 & 0.275 & --- & --- & 0.481 \\
HLE17~\cite{Verma2017_7144} & 0 & 0 & 0 & 0 & 0.092 & 0.093 & 0 \\
HLTAPW~\cite{Lehtola2021_943} & 0 & 0 & 0 & 0 & --- & --- & 0 \\
KCISK~\cite{Rey1998_581,Krieger1999_463,Krieger2001_48,Kurth1999_889,Toulouse2002_10465} & 0 & 0 & 0 & 0.012 & --- & --- & 0.041 \\
KCIS~\cite{Rey1998_581,Krieger1999_463,Krieger2001_48,Kurth1999_889,Toulouse2002_10465} & 0 & 0 & 0 & 0 & --- & --- & 0 \\
LP90~\cite{Lee1990_193} & 0 & 0.963 & 0.998 & 0.990 & 0 & 0 & 0 \\
M06-L~\cite{Zhao2006_194101,Zhao2008_215} & 0.700 & 0.661 & 0.705 & 0.696 & 0.228 & 0.181 & 0.698 \\
M11-L~\cite{Peverati2012_117} & 0.385 & 0.194 & 0.004 & 0.153 & 0.425 & 0.456 & 0.478 \\
MN12-L~\cite{Peverati2012_13171} & 0.424 & 0.266 & 0.022 & 0.227 & 0.048 & 0.086 & 0.506 \\
MN15-L~\cite{Yu2016_1280} & 0.462 & 0.184 & 3e-04 & 0.156 & 2e-04 & 0.006 & 0.594 \\
OTPSS-D~\cite{Goerigk2010_107} & 0 & 0 & 0 & 0 & 0 & 0 & 0.007 \\
PKZB~\cite{Perdew1999_2544} & 0 & 0 & 0 & 0 & 0 & 0 & 0.006 \\
R2SCANL~\cite{Mejia2020_121109,Furness2020_8208,Furness2020_9248} & 0 & 0 & 0 & 0 & 0 & 0 & 0 \\
R2SCAN~\cite{Furness2020_8208,Furness2020_9248} & 0 & 0 & 0 & 0 & 0 & 0 & 0 \\
REVM06-L~\cite{Wang2017_8487} & 0.777 & 0.702 & 0.835 & 0.783 & 5e-05 & 7e-05 & 0.767 \\
REVSCAN~\cite{Mezei2018_2469} & 0.212 & 0.063 & 0 & 0.084 & 0 & 0 & 0.288 \\
REVTM~\cite{Jana2019_6356} & 0 & 0 & 0 & 0 & 0 & 0 & 0.329 \\
REVTPSS~\cite{Perdew2009_026403,Perdew2011_179902} & 0 & 0 & 0 & 0 & 0 & 0 & 0.424 \\
RSCAN~\cite{Bartok2019_161101} & 0 & 0.008 & 0.101 & 0.160 & 0 & 0 & 0 \\
SCANL~\cite{Mejia2017_052512,Mejia2018_115161,Sun2015_036402} & 0 & 0 & 0 & 0 & 0 & 0 & 0 \\
SCAN~\cite{Sun2015_036402} & 0 & 0 & 0 & 0 & 0 & 0 & 0 \\
TM~\cite{Tao2016_073001} & 0 & 0 & 0 & 0 & 0 & 0 & 7e-04 \\
TPSSLOC~\cite{Constantin2012_035130} & 0 & 0 & 0 & 0.003 & --- & --- & 0.280 \\
TPSSLYP1W~\cite{Dahlke2005_15677} & 0.408 & 0.179 & 0 & 0.138 & 0 & 0 & 0.420 \\
TPSS~\cite{Tao2003_146401,Perdew2004_6898} & 0 & 0 & 0 & 0 & 0 & 0 & 0.008 \\
VSXC~\cite{VanVoorhis1998_400} & 0.298 & 0.084 & 0.190 & 0.260 & --- & --- & 0.294 \\
ZLP~\cite{Zhao1993_918} & 0 & 0.943 & 0.998 & 0.981 & 0.343 & 0.285 & 0 \\
\bottomrule
\end{tabular}
\end{table*}

\begin{table*}
\caption{Hybrid GGA functionals: numerical assessment of corresponding local conditions.}
\begin{tabular}{|l|c|c|c|c|c|c|c|}
\toprule
 & \makecell[c]{$E\C[n]$ \\ non-positivity} & \makecell[c]{$E\C[n\g]$ uniform \\ scaling inequality} & \makecell[c]{$T\C[n]$ \\ upper bound} & \makecell[c]{concavity of \\ $\lambda^2 E\c[n_{1/\lambda}]$} & \makecell[c]{LO extension \\ to $E\xc$} & \makecell[c]{LO} & \makecell[c]{conjecture: \\ $T_c \leq -E_c$} \\
\midrule
APBE0~\cite{Fabiano2015_122} & 0 & 0 & 0 & 0 & 0 & 0 & 0.004 \\
APF~\cite{Austin2012_4989} & 0 & 0 & 0 & 0 & 0 & 0 & 0 \\
B1LYP~\cite{Adamo1997_242} & 0.576 & 0.218 & 0.003 & 0.203 & 0 & 0 & 0.511 \\
B1PW91~\cite{Adamo1997_242} & 0 & 0 & 0 & 0 & 0 & 0 & 0 \\
B1WC~\cite{Bilc2008_165107} & 0 & 0 & 0 & 0 & 0 & 0 & 0.005 \\
B3LYP-MCM1~\cite{Caldeira2019_62} & 0.753 & 0.282 & 0.681 & 0.455 & 0 & 0 & 0.590 \\
B3LYP-MCM2~\cite{Caldeira2019_62} & 0.558 & 0.210 & 4e-04 & 0.187 & 0 & 0 & 0.504 \\
B3LYP3~\cite{Stephens1994_11623} & 0.457 & 0.174 & 0 & 0.154 & 0 & 0 & 0.451 \\
B3LYP5~\cite{Stephens1994_11623} & 0.457 & 0.174 & 0 & 0.154 & 0 & 0 & 0.451 \\
B3LYPS~\cite{Reiher2001_48} & 0.414 & 0.159 & 0 & 0.137 & 0 & 5e-05 & 0.442 \\
B3LYP~\cite{Stephens1994_11623} & 0.414 & 0.159 & 0 & 0.137 & 0 & 8e-05 & 0.442 \\
B3P86~\cite{gaussianimplementation} & 0.074 & 0.141 & 0 & 0.209 & 0 & 0 & 0.265 \\
B3PW91~\cite{Becke1993_5648} & 0 & 0 & 0 & 0 & 0 & 0 & 0 \\
B5050LYP~\cite{Shao2003_4807} & 0.457 & 0.174 & 0 & 0.154 & 0 & 0 & 0.451 \\
B97-1P~\cite{Cohen2000_160} & 0.666 & 0.463 & 0.634 & 0.649 & 0.056 & 0.042 & 0.663 \\
B97-1~\cite{Hamprecht1998_6264} & 0.698 & 0.476 & 0.686 & 0.687 & 0.002 & 0.001 & 0.693 \\
B97-2~\cite{Wilson2001_9233} & 0.665 & 0.479 & 0.605 & 0.643 & 0.065 & 0.048 & 0.666 \\
B97-3~\cite{Keal2005_121103} & 0.612 & 0.464 & 0.601 & 0.601 & 0.089 & 0.050 & 0.608 \\
B97-K~\cite{Boese2004_3405} & 0.330 & 0.004 & 0.520 & 0.342 & 0 & 0 & 0.291 \\
B97~\cite{Becke1997_8554} & 0.645 & 0.359 & 0.612 & 0.626 & 0.003 & 0.003 & 0.641 \\
BHANDHLYP~\cite{Becke1993_1372,gaussianimplementation} & 0.576 & 0.218 & 0.003 & 0.203 & 0 & 0 & 0.511 \\
BHANDH~\cite{Becke1993_1372,gaussianimplementation} & 0.576 & 0.218 & 0.003 & 0.203 & 0 & 0 & 0.511 \\
BLYP35~\cite{Renz2009_16292,Kaupp2011_16973} & 0.576 & 0.218 & 0.003 & 0.203 & 0 & 0 & 0.511 \\
CAP0~\cite{Carmona2016_120} & 0 & 0 & 0 & 0 & 0.080 & 0.081 & 0.005 \\
CASE21~\cite{Sparrow2022_6896} & 0 & 0 & 0 & 0 & 0 & 0 & 0.259 \\
EDF2~\cite{Lin2004_365} & 0.362 & 0.132 & 0 & 0.125 & 0 & 0 & 0.398 \\
HAPBE~\cite{Fabiano2015_122} & 0 & 0 & 0 & 0 & 0 & 0 & 0.012 \\
HFLYP~\cite{Lee1988_785,Miehlich1989_200} & 0.576 & 0.218 & 0.003 & 0.203 & 0 & 0 & 0.511 \\
HPBEINT~\cite{Fabiano2013_673} & 0 & 0 & 0 & 0 & 0 & 0 & 0.005 \\
KMLYP~\cite{Kang2001_11040} & 0.121 & 0.052 & 0 & 0.054 & 0 & 0 & 0.279 \\
MB3LYP-RC04~\cite{Tognetti2007_381} & 0.359 & 0.143 & 0.001 & 0.126 & 0 & 1e-04 & 0.375 \\
MPW1K~\cite{Lynch2000_4811} & 0 & 0 & 0 & 0 & 0 & 0 & 0 \\
MPW1LYP~\cite{Adamo1998_664} & 0.576 & 0.218 & 0.003 & 0.203 & 0 & 0 & 0.511 \\
MPW1PBE~\cite{Adamo1998_664} & 0 & 0 & 0 & 0 & 0 & 0 & 0.005 \\
MPW1PW~\cite{Adamo1998_664} & 0 & 0 & 0 & 0 & 0 & 0 & 0 \\
MPW3LYP~\cite{Zhao2004_6908} & 0.462 & 0.178 & 0 & 0.153 & 0 & 0 & 0.465 \\
MPW3PW~\cite{Adamo1998_664} & 0 & 0 & 0 & 0 & 0 & 0 & 0 \\
MPWLYP1M~\cite{Schultz2005_11127} & 0.576 & 0.218 & 0.003 & 0.203 & 3e-04 & 3e-04 & 0.511 \\
O3LYP~\cite{Hoe2001_319,Cohen2001_607} & 0.457 & 0.174 & 0 & 0.154 & 0.017 & 0.028 & 0.451 \\
PBE-2X~\cite{Tahchieva2018_4806} & 0 & 0 & 0 & 0 & 0 & 0 & 0.005 \\
PBE-MOL0~\cite{delCampo2012_104108} & 0 & 0 & 0 & 0 & 0 & 0 & 0.004 \\
PBE-MOLB0~\cite{delCampo2012_104108} & 0 & 0 & 0 & 0 & 0 & 0 & 0.005 \\
PBE-SOL0~\cite{delCampo2012_104108} & 0 & 0 & 0 & 0 & 0 & 0 & 0.006 \\
PBE0-13~\cite{Cortona2012_086101} & 0 & 0 & 0 & 0 & 0 & 0 & 0.005 \\
PBE38~\cite{Grimme2010_154104} & 0 & 0 & 0 & 0 & 0 & 0 & 0.005 \\
PBE50~\cite{Bernard2012_204103} & 0 & 0 & 0 & 0 & 0 & 0 & 0.005 \\
PBEB0~\cite{delCampo2012_104108} & 0 & 0 & 0 & 0 & 0 & 0 & 0.005 \\
PBEH~\cite{Adamo1999_6158,Ernzerhof1999_5029} & 0 & 0 & 0 & 0 & 0 & 0 & 0.005 \\
QTP17~\cite{Jin2018_064111} & 0.407 & 0.156 & 0 & 0.135 & 0 & 0 & 0.439 \\
REVB3LYP~\cite{Lu2013_64} & 0.438 & 0.168 & 0 & 0.145 & 0 & 0 & 0.453 \\
SB98-1A~\cite{Schmider1998_9624} & 0.803 & 0.435 & 0.903 & 0.809 & 0 & 0 & 0.786 \\
SB98-1B~\cite{Schmider1998_9624} & 0.267 & 0.049 & 0.055 & 0.215 & 0.013 & 0.019 & 0.278 \\
SB98-1C~\cite{Schmider1998_9624} & 0.635 & 0.363 & 0.593 & 0.615 & 0.004 & 0.004 & 0.633 \\
SB98-2A~\cite{Schmider1998_9624} & 0 & 0 & 0 & 0 & 0.002 & 0.012 & 0 \\
SB98-2B~\cite{Schmider1998_9624} & 0.478 & 0.180 & 0.327 & 0.432 & 2e-05 & 4e-04 & 0.484 \\
SB98-2C~\cite{Schmider1998_9624} & 0.632 & 0.371 & 0.586 & 0.611 & 0.001 & 0.002 & 0.629 \\
SOGGA11-X~\cite{Peverati2011_191102} & 0 & 0.042 & 0.176 & 0.169 & 0 & 0.003 & 0 \\
WC04~\cite{Wiitala2006_1085} & 0 & 0 & 0 & 0 & 0.123 & 0.240 & 0 \\
WP04~\cite{Wiitala2006_1085} & 0 & 0 & 0 & 0 & 0.349 & 0.468 & 0 \\
X3LYP~\cite{Xu2004_2673} & 0.462 & 0.178 & 0 & 0.153 & 0 & 0 & 0.465 \\
\bottomrule
\end{tabular}
\end{table*}

\begin{table*}
\caption{Hybrid MGGA functionals: numerical assessment of corresponding local conditions.}
\begin{tabular}{|l|c|c|c|c|c|c|c|}
\toprule
 & \makecell[c]{$E\C[n]$ \\ non-positivity} & \makecell[c]{$E\C[n\g]$ uniform \\ scaling inequality} & \makecell[c]{$T\C[n]$ \\ upper bound} & \makecell[c]{concavity of \\ $\lambda^2 E\c[n_{1/\lambda}]$} & \makecell[c]{LO extension \\ to $E\xc$} & \makecell[c]{LO} & \makecell[c]{conjecture: \\ $T_c \leq -E_c$} \\
\midrule
B0KCIS~\cite{Toulouse2002_10465} & 0 & 0 & 0 & 0 & 0 & 0 & 0 \\
B86B95~\cite{Becke1996_1040} & 0 & 0 & 0 & 0 & 0 & 9e-04 & 0 \\
B88B95~\cite{Becke1996_1040} & 0 & 0 & 0 & 0 & 0 & 9e-04 & 0 \\
B98~\cite{Becke1998_2092} & 0.202 & 0.004 & 0.145 & 0.119 & 0 & 0 & 0.172 \\
BB1K~\cite{Zhao2004_2715} & 0 & 0 & 0 & 0 & 0 & 1e-03 & 0 \\
BR3P86~\cite{Neumann1995_381} & 0.290 & 0.336 & 0.005 & 0.247 & 0 & 6e-04 & 0.305 \\
DLDF~\cite{Pernal2009_263201} & 0.543 & 0.455 & 0.523 & 0.532 & 0.298 & 0.241 & 0.540 \\
EDMGGAH~\cite{Tao2002_2335} & 0.351 & 0.166 & 0.254 & 0.276 & 0.015 & 0.013 & 0.481 \\
M05-2X~\cite{Zhao2006_364} & 0.745 & 0.696 & 0.760 & 0.743 & 0.070 & 0.042 & 0.741 \\
M05~\cite{Zhao2005_161103} & 0.748 & 0.559 & 0.767 & 0.729 & 0.075 & 0.073 & 0.731 \\
M06-2X~\cite{Zhao2008_215} & 0.681 & 0.613 & 0.720 & 0.682 & 0.058 & 0.036 & 0.674 \\
M06-HF~\cite{Zhao2006_13126} & 0.298 & 0.232 & 0.382 & 0.313 & 0.049 & 0.105 & 0.284 \\
M06~\cite{Zhao2008_215} & 0.419 & 0.229 & 0.257 & 0.375 & 0.329 & 0.330 & 0.423 \\
M08-HX~\cite{Zhao2008_1849} & 0.039 & 0.008 & 0 & 0.007 & 0.070 & 0.259 & 0.136 \\
M08-SO~\cite{Zhao2008_1849} & 0.079 & 0.019 & 0 & 0.015 & 0.149 & 0.260 & 0.196 \\
MN15~\cite{Yu2016_5032} & 0.785 & 0.612 & 0.396 & 0.631 & 0.288 & 0.276 & 0.860 \\
MPW1B95~\cite{Zhao2004_6908} & 0 & 0 & 0 & 0 & 0 & 9e-04 & 0 \\
MPW1KCIS~\cite{Zhao2005_2012} & 0 & 0 & 0 & 0 & 0 & 0 & 0 \\
MPWB1K~\cite{Zhao2004_6908} & 0 & 0 & 0 & 0 & 0 & 1e-03 & 0 \\
MPWKCIS1K~\cite{Zhao2005_2012} & 0 & 0 & 0 & 0 & 0 & 0 & 0 \\
PBE1KCIS~\cite{Zhao2005_415} & 0 & 0 & 0 & 0 & 0 & 0 & 0 \\
PW6B95~\cite{Zhao2005_5656} & 0 & 0 & 0 & 0 & 0 & 9e-04 & 0 \\
PW86B95~\cite{Becke1996_1040} & 0 & 0 & 0 & 0 & 0 & 9e-04 & 0 \\
PWB6K~\cite{Zhao2005_5656} & 0 & 0 & 0 & 0 & 0 & 1e-03 & 0 \\
REVM06~\cite{Wang2018_10257} & 0.676 & 0.591 & 0.685 & 0.670 & 0.003 & 0.002 & 0.670 \\
REVTPSSH~\cite{Csonka2010_3688} & 0 & 0 & 0 & 0 & 0 & 0 & 0.424 \\
TPSS0~\cite{Grimme2005_3067} & 0 & 0 & 0 & 0 & 0 & 0 & 0.008 \\
TPSS1KCIS~\cite{Zhao2005_43} & 0 & 0 & 0 & 0 & 0 & 0 & 0 \\
TPSSH~\cite{Staroverov2003_12129} & 0 & 0 & 0 & 0 & 0 & 0 & 0.008 \\
X1B95~\cite{Zhao2004_6908} & 0 & 0 & 0 & 0 & 0 & 0 & 0 \\
XB1K~\cite{Zhao2004_6908} & 0 & 0 & 0 & 0 & 0 & 4e-05 & 0 \\
\bottomrule
\end{tabular}
\end{table*}

\clearpage

\bibliographystyle{apsrev4-2}
\typeout{} 
\bibliography{Master,ryan_master,supp_tables/comprehensive_search}

\label{page:end}